\documentclass[11pt,a4paper]{article}

\usepackage[utf8]{inputenc}
\usepackage[T1]{fontenc}
\usepackage{amsmath, amssymb, amsthm} 
\usepackage{graphicx}                 
\usepackage{geometry}                 
\usepackage{authblk}                  
\usepackage{cite}                     
\usepackage{hyperref}                 
\hypersetup{
    colorlinks=true,
    linkcolor=blue,
    filecolor=magenta,      
    urlcolor=cyan,
    citecolor=blue,
}
\usepackage{natbib}

\usepackage{lipsum}

\usepackage{multirow}
\usepackage{algorithm}
\usepackage{algpseudocode}
\usepackage{enumitem}

\newtheorem{prop}{Proposition}
\newtheorem{theorem}{Theorem}

\title{Multivariate Multinomial Logit Model with ANOVA Decomposition for Correlated Categorical Outcomes}

\author[1]{Sohyeon Kim\thanks{Corresponding author: skim226@ncsu.edu}}
\author[1]{Luo Xiao}
\author[2]{Xinming An}
\author[1]{Wenbin Lu}

\affil[1]{Department of Statistics, North Carolina State University, Raleigh}
\affil[2]{Department of Anesthesiology, University of North Carolina at Chapel Hill, Chapel Hill}

\date{June 12, 2026}

\begin{document}

\maketitle

\begin{abstract}
In medical research, patients often have multiple interdependent outcomes, such as posttraumatic stress disorder (PTSD), depression, and pain among trauma survivors. Most existing research uses multinomial regression to analyze these interdependent outcomes separately, which ignores correlations between concurrent conditions. This omission may lead to loss of information and reduced predictive accuracy. Accounting for correlations between multiple categorical outcomes requires a high-dimensional parameter space, making model estimation challenging. In this paper, we propose a multivariate multinomial logit model that captures outcome correlations and uses the ANOVA decomposition of the parameter space to reduce the number of parameters. The ANOVA decomposition enables explicit conditional model formulations, which allow for a computationally much simpler composite likelihood for model estimation. We develop an efficient Minorization-Maximization (MM) algorithm to maximize the composite likelihood, which also incorporates variable selection via a bridge penalty. Simulation studies are conducted to evaluate our method, demonstrating its accuracy in parameter estimation and variable selection. We further illustrate our method using data from the AURORA study.
\end{abstract}

\section{Introduction}
Comorbidities are common in healthcare, where patients often develop multiple co-occurring medical conditions \citep{valderas2009defining}. This phenomenon is particularly true among trauma survivors, who frequently experience a wide range of disorders, such as pain, PTSD, depression, and somatic symptoms, with overlapping symptoms \citep{mclean2020aurora}. Although the underlying biological mechanism and risk factors may be shared by these disorders, clinical care remains largely organized around disorder-specific frameworks. As a result, patients are typically treated in parallel by different clinical specialties, each targeting a single diagnosis, leading to fragmented care. To address this issue, integrative and transdiagnostic approaches that focus on multiple disorders are increasingly recognized as an effective way to disentangle shared mechanisms and risk factors \citep{dalgleish2020transdiagnostic, gutner2016emergence, gutner2019dealing}. However, the intercorrelation between these multiple disorders presents a significant analytical challenge that standard statistical models are not equipped to handle. 

As a foundational framework to capture interdependencies between outcomes, the Ising model \citep{cheng2014sparse} and its variants have been widely utilized. However, these models are fundamentally restricted to binary outcomes. While the multivariate multinomial logistic framework serves as the natural extension for categorical data, capturing the interdependencies among multiple multinomial outcomes inevitably introduces a severe parameter dimensionality problem. For instance, attempting to model all possible combinations of outcomes leads directly to a combinatorial explosion in the number of parameters, which makes it computationally infeasible \citep{amemiya1978two, ben1985discrete}. Also, model parameters in such a full model are often difficult to interpret.
Multinomial probit models \citep{hausman1978conditional} and mixed logit models \citep{hensher2003mixed} can capture the dependencies between multinomial outcomes theoretically, but they are frequently hampered by a heavy computational burden \citep{geweke1994alternative, geweke1997statistical}.

The computational challenges associated with models for multivariate multinomial outcomes become more severe when the number of predictors is large or even high-dimensional, rendering variable selection for such models almost infeasible. In contrast, various variable selection methods for multivariate continuous outcomes have been developed in the literature; see, e.g., 
\cite{yuan2007dimension}, \cite{chen2013reduced}, and \cite{ma2014learning}. For multinomial models, however, recent literature has begun to address variable selection specifically for univariate outcomes. Because a single predictor in a multinomial model involves multiple category-specific parameters, variable-level selection is inherently required rather than individual parameter selection. To address this grouped structure, \cite{zahid2013multinomial, tutz2015variable} develops a group-penalization approach that selects entire variables rather than individual parameters, effectively accommodating both global and category-specific predictors. In addition, \cite{chen2013variable} introduces a penalized Dirichlet-multinomial regression using a sparse group $\ell_1$ penalty to achieve both group-level and within-group variable selection. While these methods effectively handle univariate multinomial outcomes, extending them to multivariate multinomial responses remains computationally challenging due to the complex outcome correlations. To address this, we generalized the multivariate multinomial logit model of \cite{bel2014multivariate} using the ANOVA decomposition to capture the outcome correlations and reduce the dimension of the parameter space.
This decomposition allows for an explicit multinomial logit formulation of the conditional model for each categorical outcome, 
which provides improved interpretability of the parameters. Moreover,
conditional models enable model estimation by the composite likelihood approach, a computationally much less intensive strategy. Based on composite likelihood, we developed a computationally efficient algorithm to perform variable selection for multivariate multinomial models, using the MM algorithm.
Importantly, a key strength of the proposed model lies in its ability to produce a robust solution even when certain combinations of outcomes are absent from the data. We also conducted simulation studies to test the limits of this robustness by evaluating how the model performs with missing combinations under different settings.

Section \ref{Sec: Model} details the model, and Section \ref{sec: Methods} describes the methodology, including the design matrix, penalty, algorithm, and selection of the tuning parameters. We evaluate the performance of the proposed method using simulations in Section \ref{sec: simulation} and apply it to real data from a large-scale study of trauma survivors in Section \ref{sec: real data analysis}.
We conclude the paper with a discussion in Section \ref{sec:discussion}.

\section{Model} \label{Sec: Model}

In this section, we discuss the model specification for the multivariate multinomial logit model. We first consider the model specification for binary outcomes in Section \ref{sec: multivariate binary model} and then expand the binary case to multinomial outcomes in Section \ref{sec: multivariate multinomial model}.

\subsection{Multivariate Binary Model} \label{sec: multivariate binary model}

We first consider a multivariate logit model for correlated binary outcomes, following the approach of \cite{russell2000analysis}. For each individual $i$ (from 1 to $n$), let the outcome $Y _i$ be a $K$-dimensional random vector of binary choices and $Y_{ik} \in \{0, 1\}$ is the choice for the $k$-th outcome. This results in $2^K$ possible realizations for the random vector $Y_i$.

To account for potential correlations among the $K$ choices, the model is defined by the conditional probability of each outcome $Y_{ik}$, given an individual's characteristics $X_i$, and their other choices $y_{i\ell}$ (where $\ell \ne k$). This conditional probability is defined using a logit function:
\begin{equation}\label{eq: cond_prob}
    P(Y_{ik} = 1|y_{i\ell} \text{ for }\ell \ne k, X_i) = \frac{\text{exp}(Z_{ik})}{1 + \text{exp}(Z_{ik})},
\end{equation}
where the linear predictor $Z_{ik}$ is given by
\begin{equation} \label{eq: Model}
    Z_{ik} = \alpha_k + X_i\beta_k + \sum_{\ell \ne k}I(y_{i\ell} = 1)(\psi_{k\ell} + X_i\delta_{k\ell}),
\end{equation}
where $X_i$ is a $p$-dimensional row vector, and $\beta_k$ and $\delta_{kl}$ are $p$-dimensional column vectors.
To ensure an identifiable model, we set category `0' as the reference level by fixing its corresponding parameters to zero. Consequently, the parameter in Equation \eqref{eq: Model} represents the effect for category `1' relative to this reference, where $\alpha_k$ is the baseline intercept for outcome $k$, $\beta_k$ is the vector of coefficients corresponding to the covariates $X_i$, and $\psi_{k\ell}$ and $\delta_{k\ell}$ are the association parameters that capture the dependency of outcome $k$ on outcome $\ell$. 
The term $(\psi_{k\ell} + X_i\delta_{k\ell})$ quantifies the relationship between outcomes $Y_{ik}$ and $Y_{i\ell}$. Specifically, a positive value indicates that it is more likely to have $Y_{ik} = 1$ when $Y_{i\ell} = 1$ than $Y_{i\ell} = 0$. On the other hand, a value of zero implies that these two outcomes are conditionally independent. 
The association term $(\psi_{k\ell} + X_i\delta_{k\ell})$ represents the log odds ratio for the pair of outcomes $(k, \ell)$. This association holds regardless of the values of the other outcomes $(Y_{im}, m \ne k,\ell)$, as their respective parameters factor out and cancel during the log odds ratio calculation. Equation \eqref{eq:odds ratio} illustrates this property using the simplest case where all other outcomes take the reference `0' level:
\begin{equation}\label{eq:odds ratio}
    \text{ln}\Bigg(\frac{P(0, \dots, 0, y_k, 0, \dots, 0, y_\ell, 0, \dots, 0|X_i)P(0, \dots, 0|X_i)}{P(0, \dots, 0, y_k, 0, \dots, 0|X_i)P(0, \dots, 0, y_{\ell}, 0, \dots, 0)|X_i)}\Bigg) = \psi_{k\ell} + X_i\delta_{k\ell}.
\end{equation}

\subsection{Multivariate Multinomial Model} \label{sec: multivariate multinomial model}
We now extend the binary framework to the multivariate multinomial model. We consider $K$ multinomial outcomes, where the $k$-th outcome for an individual $i$, denoted as $Y_{ik}$, can take one of $J_k$ possible categories (i.e., $Y_{ik} \in \{1, \dots, J_k\}$). The total number of possible outcome combinations for the vector $Y_i$ is $\prod_{k=1}^K J_k$.

The conditional probability that the $k$-th outcome is category $j$, given the individual's covariates $X_i$ and their other choices $y_{i\ell}$ ($\ell \ne k$), is modeled using the softmax function:
\begin{equation}\label{eq: conditional prob multinom}
    P(Y_{ik} = j|y_{i\ell} \text{ for }\ell \ne k, X_i) = \frac{\exp(Z_{ik,j})}{\sum_{h = 1}^{J_k}\exp(Z_{ik,h})},
\end{equation}
where the linear predictor $Z_{ik,j}$ is defined as:
\begin{equation} \label{eq: mml}
    Z_{ik,j} = \alpha_{k,j} + X_i\beta_{k, j} + \sum_{\ell \ne k}(\psi_{k\ell, jy_{i\ell}} + X_i \delta_{k\ell, jy_{i\ell}}),
\end{equation}
where $\beta_{k,j}$ and $\delta_{kl,jh}$ are $p$-dimensional column vectors.
In this formulation, the parameters $\alpha_{k,j}$ and $\beta_{k,j}$ are the category-specific main effects, representing intercepts and covariate effects for category $j$ of outcome $k$. The dependency between choices is captured by the association parameters $\psi_{k\ell, jy_{i\ell}}$ and $\delta_{k\ell, jy_{i\ell}}$, which quantify the association between category $j$ for outcome $k$ and category $y_{i\ell}$ for outcome $\ell$. This association term is equivalent to a generalized log-odds ratio.

Having specified the conditional models in Equations \eqref{eq: conditional prob multinom} and \eqref{eq: mml}, we now demonstrate how they serve as a strategic foundation to derive the joint probability, which is the ultimate goal of the MML framework. The joint probability for a specific outcome vector $y_i = (y_{i1}, \dots, y_{iK})$ is given by:
\begin{equation}\label{eq: joint_prob}
    P(Y_i = y_i|X_i) = \frac{\exp(\mu_{y_i})}{\sum_{s \in S}\exp(\mu_{s})},
\end{equation}
where $S$ is the set of all possible outcome vectors and the potential function $\mu_{y_i}$ can be interpreted as a sum of the main effects and pairwise interaction terms:
\begin{equation} \label{eq: joint mu}
    \mu_{y_i} = \sum_{k = 1}^K(\alpha_{k,y_{ik}} + X_i\beta_{k,y_{ik}}) + \sum_{k < \ell} (\psi_{k\ell, y_{ik}y_{i\ell}} + X_i\delta_{k\ell, y_{ik}y_{i\ell}}).
\end{equation}
Specifically, the ANOVA decomposition provides a connection that links the joint probability to the conditional probability. From the Equation \eqref{eq: conditional prob multinom} and \eqref{eq: mml}, the joint probability (Equation \eqref{eq: joint_prob} and \eqref{eq: joint mu}) can be derived, and conversely, the equation regarding joint distribution recovers the conditional probability. The proofs are provided in the Appendix \ref{appendix: connection between conditional and joint probability}.

The model defined by $\mu_{y_i}$ is overparameterized, so several constraints are required to ensure the identifiability of the parameters. First, we apply the standard multinomial logit restriction by setting the first category as the reference for all main effects, which fixes $\alpha_{k,1} = 0$ and $\beta_{k,1} = \mathbf{0}$ for all $k$. 
Furthermore, the log-odds ratio for the pair of outcomes $(k, \ell)$ in a multivariate multinomial model can be written as follows:
\begin{align}\label{eq:odds ratio multi}
    \text{ln}\Bigg(\frac{P(1, \dots, 1, y_{ik}, 1, \dots, 1, y_{i\ell}, 1, \dots, 1|X_i)P(1, \dots, 1|X_i)}{P(1, \dots, 1, y_{ik}, 1, \dots, 1|X_i)P(1, \dots, 1, y_{i\ell}, 1, \dots, 1)|X_i)}\Bigg) \\= \psi_{k\ell,y_{ik}y_{i\ell}} + X_i\delta_{k\ell,y_{ik}y_{i\ell}}. \nonumber
\end{align}
Note that when we swap outcomes $y_{ik}$ and $y_{il}$ on the left-hand side of the above equation, it will give the same log-odds ratio. In other words, the right-hand side of Equation \eqref{eq:odds ratio multi} is equivalent to $\psi_{\ell k,y_{i\ell}y_{ik}} + X_i\delta_{\ell k,y_{i\ell}y_{ik}}$, which leads to the following symmetry constraints: $\psi_{k\ell, jh} = \psi_{\ell k, hj}$ and ${\delta}_{k\ell, jh} = {\delta}_{\ell k, hj}$. Lastly, to ensure full identifiability, any association term involving a reference category is set to zero, which means $\psi_{k\ell, j1} = \psi_{k\ell, 1h} = 0$ and ${\delta}_{k\ell, j1} = {\delta}_{k\ell, 1h} = \mathbf{0}$. With these constraints, the number of free intercept parameters for the main effects ($\alpha_{k,j}$) is $\sum_{k=1}^K (J_k - 1)$, and for the association effects ($\psi_{k\ell, jh}$), it is $\sum_{k < \ell}(J_k - 1)(J_\ell - 1)$. 
Moreover, Equation \eqref{eq:odds ratio multi} also implies that the proposed model is estimable even if not all combinations of outcomes are observed in the data.

\begin{prop} \label{prop: identifiability missing combi}
    The proposed model is estimable as long as the data contain observations of the form $(1, \dots, j, \dots, 1)^T$ or $(1, \dots, j, 1, \dots, h, \dots, 1)^T$ for all relevant categories $j$ and $h$, where `1' represents the reference level.  
\end{prop}

\section{Methods} \label{sec: Methods}

Parameter estimation for the multivariate multinomial logit model is typically achieved by maximizing the log-likelihood function. The standard approach, proposed by \cite{russell2000analysis}, is based on the joint probability distribution in Equation \eqref{eq: joint_prob}. Although this method produces statistically efficient estimators, it becomes computationally intractable as the number of outcomes or categories grows. The primary challenge is the required summation over the entire outcome space, which has a size of $\prod_{k=1}^K J_k$ and expands exponentially, making the computation infeasible when the number of outcomes is large.

To overcome this computational burden, we adopt a composite likelihood approach, following the work of \cite{bel2014multivariate}. This method constructs a pseudo-likelihood using the product of the simpler conditional probabilities (Equation \eqref{eq: conditional prob multinom}) instead of the joint probability. This composite conditional likelihood (CCL) approach significantly reduces computational complexity by avoiding the summation over the entire outcome space. As established in the literature \citep{lindsay11988composite, varin2011overview}, the estimators derived from maximizing a composite likelihood are consistent, although they may not be as efficient as the maximum likelihood estimators.

The composite log-likelihood function for our model is defined as $$\ell^c(\theta;y) 
    = \\ \sum_{i=1}^n\sum_{k=1}^K\sum_{j=1}^{J_k}I(Y_{ik} = j)\log\big\{P(Y_{ik} = j|y_{i\ell} \text{ for } \ell \ne k, X_i)\big\}.$$ This expression can be rewritten as:
\begin{equation} \label{eq: CCL}
    \ell^c(\theta;y) = \sum_{i=1}^n\sum_{k=1}^K\sum_{j=1}^{J_k}I(Y_{ik} = j)\bigg\{Z_{ik,j} - \log\bigg(\sum_{h=1}^{J_k} \exp(Z_{ik,h})\bigg)\bigg\}.
\end{equation}

\subsection{Matrix Formulation}

To express the objective function of the model in a compact matrix form suitable for model estimation, we construct the overall design matrix, $\tilde{\mathbf{X}}$. This process begins by vectorizing all the model parameters. 
\begin{align*}
    &\boldsymbol{\alpha}_k = (\alpha_{k,1}, 
    \dots, \alpha_{k,J_k})^T, \; \boldsymbol{\alpha} = (\boldsymbol{\alpha}_1^T, \dots, \boldsymbol{\alpha}_K^T)^T,\\
    &\boldsymbol{\beta}_{k} = (\beta_{k,1}^T, \dots,\beta_{k,J_k}^T)^T  \; \boldsymbol{\beta} = (\boldsymbol{\beta}_1^T, \dots, \boldsymbol{\beta}_K^T)^
    T,\\
    &\boldsymbol{\psi}_{k\ell} = (\psi_{k\ell,11}, \psi_{k\ell, 12}, \dots, \psi_{k\ell,J_kJ_\ell})^T,\; \\
    &\boldsymbol{\psi}_k = (\boldsymbol{\psi}_{1k}^T, \dots, \boldsymbol{\psi}_{(k-1)k}^T, \boldsymbol{\psi}_{k(k+1)}^T, \dots\boldsymbol{\psi}_{kK}^T)^T,\\
    &\boldsymbol{\delta}_{k\ell} = (\delta_{k\ell,11}^T, \delta_{k\ell, 12}^T, \dots, \delta_{k\ell,J_kJ_\ell}^T)^T,\; \\
    &\boldsymbol{\delta}_k = (\boldsymbol{\delta}_{1k}^T, \dots, \boldsymbol{\delta}_{(k-1)k}^T, \boldsymbol{\delta}_{k(k+1)}^T, \dots\boldsymbol{\delta}_{kK}^T)^T,\\
    &\boldsymbol{\psi} = (\boldsymbol{\psi}_{12}^T, , \boldsymbol{\psi}_{13}^T,\dots, \boldsymbol{\psi}_{(K-1)K}^T)^T, \;\boldsymbol{\delta} = (\boldsymbol{\delta}_{12}^T, , \boldsymbol{\delta}_{13}^T,\dots, \boldsymbol{\delta}_{(K-1)K}^T)^T.
\end{align*}
Due to the identifiability constraints, all parameters corresponding to the reference level are fixed at zero. We therefore define the vector of free parameters as:
\begin{align*}
    &\tilde{\boldsymbol{\alpha}}_k = (\alpha_{k,2}, 
    \dots, \alpha_{k,J_k})^T, \; \tilde{\boldsymbol{\alpha}} = (\tilde{\boldsymbol{\alpha}}_1^T, \dots, \tilde{\boldsymbol{\alpha}}_K^T)^T,\\
    &\tilde{\boldsymbol{\beta}}_{k} = (\beta_{k,2}^T, \dots,\beta_{k,J_k}^T)^T  \; \tilde{\boldsymbol{\beta}} = (\tilde{\boldsymbol{\beta}}_1^T, \dots, \tilde{\boldsymbol{\beta}}_K^T)^
    T,\\
    &\tilde{\boldsymbol{\psi}}_{k\ell} = (\psi_{k\ell,22}, \psi_{k\ell, 23}, \dots, \psi_{k\ell,J_kJ_\ell})^T,\;\\
    &\tilde{\boldsymbol{\psi}}_k = (\tilde{\boldsymbol{\psi}}_{1k}^T, \dots, \tilde{\boldsymbol{\psi}}_{(k-1)k}^T, \tilde{\boldsymbol{\psi}}_{k(k+1)}^T, \dots, \tilde{\boldsymbol{\psi}}_{k,K}^T)^T,\\
    &\tilde{\boldsymbol{\delta}}_{k\ell} = (\delta_{k\ell,22}^T, \delta_{k\ell, 23}^T, \dots, \delta_{k\ell,J_kJ_\ell}^T)^T,\;\\
    & \tilde{\boldsymbol{\delta}}_k = (\tilde{\boldsymbol{\delta}}_{1k}^T, \dots,  \tilde{\boldsymbol{\delta}}_{(k-1)k}^T, \tilde{\boldsymbol{\delta}}_{k(k+1)}^T\dots, \tilde{\boldsymbol{\delta}}_{(K-1)K}^T)^T,\\
    &\tilde{\boldsymbol{\psi}} = (\tilde{\boldsymbol{\psi}}_{12}^T, \tilde{\boldsymbol{\psi}}_{13}^T, \dots, \tilde{\boldsymbol{\psi}}_{(K-1)K}^T)^T, \; \tilde{\boldsymbol{\delta}} = (\tilde{\boldsymbol{\delta}}_{12}^T, \tilde{\boldsymbol{\delta}}_{13}^T, \dots, \tilde{\boldsymbol{\delta}}_{(K-1)K}^T)^T.
\end{align*}
To accommodate free model parameters, we introduce transformation matrices $\mathbf{Q}_k,\mathbf{A}_k$ and $\textbf{R}$. 
The transformation matrix $\textbf{Q}_k$ is defined as follows:
\begin{equation*}
    \textbf{Q}_k = (\textbf{0}_{(J_k - 1) \times 1}, \textbf{I}_{J_k - 1})^T,
\end{equation*}
where $\textbf{0}_{d_1\times d_2}$ is the zero matrix of dimension $d_1 \times d_2$, and $\textbf{I}_d$ is the identity matrix of dimension $d \times d.$ Then, $\boldsymbol{\alpha}_k =\mathbf{Q}_k\tilde{\boldsymbol{\alpha}}_k$.
Next, for the association parameter $\boldsymbol{\psi}$, we need a selector matrix $\textbf{A}_k$, which makes $\boldsymbol{\psi}_k = \textbf{A}_k\boldsymbol{\psi}$, and a transformation matrix $\textbf{R}$, which makes $\boldsymbol{\psi} = \textbf{R}\tilde{\boldsymbol{\psi}}$ and $\boldsymbol{\psi}_{k\ell} = \textbf{R}_{k\ell}\tilde{\boldsymbol{\psi}}_{k\ell}$. Let
\begin{equation*}
    \textbf{A}_{k} = \bigg(\text{blockdiag}(\textbf{A}_{k,a})\bigg|\textbf{0}_{(J_k\sum_{\ell \ne k}J_\ell)\times(\sum_{\ell,s > k}J_\ell J_s)}\bigg),
\end{equation*}
where
\begin{align*}
    \textbf{A}_{k,a} = \bigg(\textbf{0}_{J_kJ_a\times\sum_{a<\ell <k}J_\ell J_a},\textbf{I}_{J_kJ_a},\textbf{0}_{J_kJ_a\times \sum_{k < \ell\le K}J_\ell J_a}\bigg)\\
    \text{ for }a < k, \textbf{A}_{k,k} = \textbf{I}_{J_k\sum_{\ell > k}J_\ell}.
\end{align*}
For the transformation matrix $\textbf{R}$,
\begin{equation*}
    \textbf{R} = \text{blockdiag}(\textbf{R}_{12},\textbf{R}_{13},\dots,\textbf{R}_{1K},\textbf{R}_{23}\dots,\textbf{R}_{2K},\dots,\textbf{R}_{(K-1)K}),
\end{equation*}
where $\textbf{R}_{k\ell} = (\Big[\textbf{R}_{k\ell}^{(1)}\Big]^T,\Big[\textbf{R}_{k\ell}^{(2)}\Big]^T )^T$ and 
\begin{equation*}
    \textbf{R}_{k\ell}^{(1)} = \textbf{0}_{J_\ell\times (J_k-1)(J_\ell - 1)},\quad \textbf{R}_{k\ell}^{(2)} = \textbf{I}_{(J_k - 1)}\otimes \begin{pmatrix}\boldsymbol{0}_{(J_\ell - 1)\times 1}^T\\\textbf{I}_{(J_\ell - 1)}\end{pmatrix},
\end{equation*}
where $\otimes$ is a Kronecker product.

Evaluating the composite likelihood requires the linear predictor $Z_{ik,j}$ for the scenario where subject $i$'s $k$-th outcome is category $j$. To isolate the active parameter for this scenario, we define binary selection matrices: $\textbf{U}_{k,j}$ for the main effect ($\boldsymbol{\alpha}_k$) and $\textbf{W}_{k,j}$ for the association effect ($\boldsymbol{\psi}_k$). Each column in these matrices corresponds to $\boldsymbol{\alpha}_k$ and $\boldsymbol{\psi}_k$, respectively. For illustration, consider the setting with $K=3$ outcomes and $(2, 3, 2)$ categories for each outcome, respectively. Then, for $Y_i = (1, 3, 1)$, we have $U_{i1,2} = (0, 1)$ and $W_{i1,2} = (0, 0, 0, 0, 0, 1, 0, 0, 1, 0)$, where $U_{ik,j}$ and $W_{ik,j}$ are the $i$-th row vectors of $\textbf{U}_{k,j}$ and $\textbf{W}_{k,j}$, respectively. This process is repeated for all subjects (from $i = 1$ to $n$), for all outcomes (from $k = 1$ to $K$), and for all non-baseline categories (from $j = 2$ to $J_k$).

These matrices map the smaller vectors of identifiable ``free'' parameters corresponding to the $k$-th outcome, $\tilde{\boldsymbol{\alpha}}_k$ and $\tilde{\boldsymbol{\psi}}_k$ to the full, overparameterized vectors corresponding to the $k$-th outcome. This is achieved through the relationships $\boldsymbol{\alpha}_k = \mathbf{Q}_k\tilde{\boldsymbol{\alpha}}_k$ and $\boldsymbol{\psi}_k = \textbf{A}_k\mathbf{R}\tilde{\boldsymbol{\psi}}_k$. 
A similar formulation is applied to the slope parameters. Applying a  matrix operation called the Khatri-Rao product ($\odot$) with covariate $X_i^T$, we obtain the vectors of free parameters $\tilde{\boldsymbol{\beta}}$ and $\tilde{\boldsymbol{\delta}}$.
Then, these components are combined to form a modified design matrix as follows:
\begin{align*}
    \tilde{\mathbf{X}}_{ik,j} = \bigg[ {U}_{ik,j}\mathbf{Q}_{k} \mid {W}_{ik,j}\mathbf{A}_{k}\mathbf{R} \mid (({U}_{ik,j}\mathbf{Q}_{k})^T \odot {X}_i^T)^T \mid \\(({W}_{ik,j}\mathbf{A}_{k}\mathbf{R})^T \odot {X}_i^T)^T \bigg],
\end{align*}
where $U_{ik,j}$ and $W_{ik,j}$ are the $i$-th row vectors of $\textbf{U}_{k,j}$ and $\textbf{W}_{k,j}$, respectively.
Then, the linear predictor $Z_{ik,j}$ can be expressed in a simple linear form $Z_{ik,j} = \tilde{\mathbf{X}}_{ik,j}\tilde{\boldsymbol{\theta}}$, where $\tilde{\boldsymbol{\theta}}$ is the concatenated vector of all free parameters, defined as $\tilde{\boldsymbol{\theta}} = (\tilde{\boldsymbol{\alpha}}^T, \tilde{\boldsymbol{\psi}}^T, \tilde{\boldsymbol{\beta}}^T, \tilde{\boldsymbol{\delta}}^T)^T.$

\subsection{Group Bridge Penalty}
The proposed model is highly parametrized, as the inclusion of all pairwise interaction terms causes the total number of parameters to grow rapidly. To effectively manage complexity and identify the most significant interactions, we incorporate a group bridge penalty into the estimation framework. This penalty is applied exclusively to the slope coefficients associated with the covariates, namely $\boldsymbol{\beta}$ and $\boldsymbol{\delta}$. The intercept parameters, $\boldsymbol{\alpha}$ and $\boldsymbol{\psi}$, are not penalized.

The key to this approach lies in the grouping structure: all coefficients corresponding to a single covariate are consolidated into a distinct group. Since each parameter in our model relates to only one covariate, these groups are non-overlapping. The primary advantage of the group bridge penalty is its ability to perform simultaneous variable selection at two levels: it can identify important groups (i.e., influential covariates) while also selecting the most important individual coefficients within those groups \citep{huang2009group}.

Let $\theta_{A_p}$ denote the vector of coefficients associated with the $p$-th covariate, $A_p$. Then, the final objective function to be maximized is the penalized composite log-likelihood (from Equation \eqref{eq: CCL}):
\begin{align} \label{eq:gbridge}
    \mathcal{O}(\theta) = \ell^c(\theta;y) - \lambda\sum_{g = 1}^G c_g\|\theta_{A_g}\|_1^\gamma.
\end{align}
Here, $\lambda > 0$ is the primary tuning parameter that controls the overall strength of the penalty. Following the recommendation of \cite{huang2009group}, we define the group-specific weights as $c_g = |A_g|^{1-\gamma}$ to adjust for differences in group size, where $|A_g|$ is the number of coefficients in group $g$. We set the bridge exponent $\gamma = 1/2$, a value that specifically enables the penalty to perform selection at both the group and individual variable levels simultaneously.

\subsection{Asymptotic Properties} \label{subsec: asymptotic property}

An important theoretical advantage of incorporating the group bridge penalty into our proposed model is that it enjoys the oracle property in group-level variable selection. Building on the theoretical framework established by \cite{huang2009group, huang2014group}, we extend these asymptotic properties to accommodate our specific setting.  

For the theoretical proofs, it is mathematically convenient to formulate the estimation as a minimization problem. Let $L^c(\theta;y) = -\ell^c(\theta;y)$ denote the negative composite log-likelihood. The equivalent objective function to minimize is then defined as:
\begin{equation}\label{appendix eq: objective function with neg log}
    Q(\theta) = L^c(\theta;y) + \lambda\sum_{g = 1}^Gc_g\|\theta_{A_g}\|_1^\gamma.
\end{equation}
All subsequent theoretical developments are based on minimizing $Q(\theta)$.

To establish the oracle property of our proposed estimator, we require certain regularity conditions. These conditions are closely adapted from \cite{huang2009group, huang2014group} to accommodate our composite likelihood framework. We show that, for $0 < \gamma < 1$, the group bridge estimators correctly select groups of nonzero coefficients with probability converging to one under reasonable conditions. The asymptotic distributions of the estimators of the coefficients in nonzero groups are derived. 
Without loss of generality, suppose that
\begin{equation*}
    \theta_{A_g} \ne 0,\;\; 1\le g \le G_1,\quad\theta_{A_g} = 0,\;\;G_1 + 1 \le g \le G.
\end{equation*}
Let $\mathcal{T}_2 = \cup_{g = G_1 + 1}^GA_g$ be the union of the groups with zero coefficients and $\mathcal{T}_1 = \mathcal{T}_2^c$. Denote by $\theta_0$ the true value of $\theta$, write $\theta_{0\mathcal{T}_1}$ and $\theta_{0\mathcal{T}_2}$ the true value of $\theta$ with index belonging to $\mathcal{T}_1$ and $\mathcal{T}_2$, respectively.The true model is fully explained by the first $G_1$ groups since $\theta_{0\mathcal{T}_2} = 0$. We allow the number of covariates $d = d_n$ to grow at a certain rate $n > d_n \rightarrow \infty$.
The following conditions are needed:
\begin{itemize}
    \item [(C1)] $d_n^4/n \rightarrow 0.$
    \item [(C2)] If $\rho_n$ and $\rho_n^*$ are the smallest and largest eigenvalues of $H(\theta_0) = E[\nabla^2L^c(\theta_0)]$. $C_n^* = \text{max}_p\sum_{g = 1}^GI(p\in A_g)$ is bounded and 
    \begin{equation*}
        \frac{\lambda_n^2}{n}\sum_{g = 1}^{G_1}c_g^2\|\theta_{0A_g}\|_1^{2\gamma - 2}|A_g|\le d_nM_n,\quad M_n = O_p(1),
    \end{equation*}
    where the constant $c_g$'s satisfy $\underset{1\le g\le G}{\text{min}}c_g \ge1$ and $\lambda_n/(n^{\gamma/2}\rho_n^*d_n^{1-\gamma/2}) \rightarrow \infty$.
    \item[(C3)] There exists a constant $r > 0$ such that $\rho_n > r$. For fixed unknown $\{\mathcal{T}_1, \theta_{0\mathcal{T}_1},G_1\}$,
    \begin{align*}
        \frac{\lambda_n}{\sqrt{n}}\rightarrow\lambda_0,\quad \frac{1}{\rho_n} + \rho_n^* + \sum_{g = 1}^Gc_g^2 = O(1),\\
        \quad \frac{\lambda_n}{n^{\gamma/2}\rho_n^*d_n^{1-\gamma/2}}\rightarrow \infty\text{ as }n \rightarrow \infty.
    \end{align*}
\end{itemize}

\begin{theorem} \label{thm: theta distance}
    Under (C1) - (C2), 
    \begin{equation*}
        \|\hat{\theta} - \theta_0\|_2 = O_p(\sqrt{d_n/n}).
    \end{equation*}
\end{theorem}

\begin{theorem} \label{thm: consistency}
    Suppose (C1) - (C3) hold. If $\{\mathcal{T}_1, \theta_{0\mathcal{T}_1}, G_1\}$ are fixed and unknown, and $\hat{\theta}_{n\mathcal{T}_1}$ and $\hat{\theta}_{n\mathcal{T}_2}$ are the estimators of $\theta_{0\mathcal{T}_1}$ and $\theta_{0\mathcal{T}_2}$ from $\hat{\theta}$, respectively, then the followings hold.
    
    \begin{enumerate}[label=(\roman*)]
        \item $P(\hat{\theta}_{n\mathcal{T}_2} = 0)\rightarrow 1.$

        \item $\sqrt{n}(\hat{\theta}_{n\mathcal{T}_1} - \theta_{0\mathcal{T}_1})\overset{d}{\rightarrow}argmin\{V_1(a):a\in \mathbb{R}^{|\mathcal{T}_1|}\}$, where
        \begin{align*}
            V_1(a) &= a'W + \frac{1}{2}a'H_\theta a + 
            \gamma\lambda_0\sum_{g = 1}^{G_1}c_g\|\boldsymbol{\theta}_{0A_g}\|_1^{\gamma - 1}\sum_{k \in A_g\cap \mathcal{T}_1}\\&\Big\{a_k\text{sgn}(\theta_{0k})I(\theta_{0k}\ne 0) + |a_k|I(\theta_{0k} = 0)\Big\},
        \end{align*}
        with $W$ distributed as $N(0, J_{\theta_0})$, where $J_\theta = \text{var}(\nabla L_1(\theta))$. In general, the estimator of coefficients in the non-zero group is $\sqrt{n/d_n}$-consistent and converges to the argmin of the Gaussian process $V_1$.
    \end{enumerate}
\end{theorem}

The proofs of both Theorems are provided in Appendix \ref{appendix: Proof of asymptotic properties}. 

\subsection{Algorithm} \label{subsec: algorithm}

To optimize the penalized objective function in Equation \eqref{eq:gbridge}, we employ a Majorize-Minorize (MM) algorithm, following the methodology of \cite{hunter2004tutorial}. An MM algorithm is an iterative procedure that maximizes a complex objective function, $f(\theta)$, by repeatedly maximizing a simpler surrogate function, $g(\theta|\theta^{(m)})$, that minorizes it. This surrogate must satisfy the conditions 
\begin{align} \label{eq: surrogate condition}
    g(\theta|\theta^{(m)}) &\le f(\theta) \text{ for all } \theta,\\
    g(\theta^{(m)}|\theta^{(m)}) &= f(\theta^{(m)}), \nonumber
\end{align}
ensuring that each update increases the value of the true objective function. We formally derive this surrogate function in the following proposition.

\begin{prop}\label{prop:mm}

    For a given observation $i$ and outcome $k$ with category $j$, let the log-likelihood component be $f(Z_{ik,j}) = I(Y_{ik} = j)\bigg\{Z_{ik,j} - \log\bigg(\sum_{h=1}^{J_k} \exp(Z_{ik,h})\bigg)\bigg\}$. Given the linear predictors $Z_{ik,j}^{(m)}$ from iteration $m$, the function $f(Z_{ik,j})$ is minorized by the quadratic surrogate function $g(Z_{ik,j} | Z_{ik,j}^{(m)})$, defined as:
\begin{align*}
     g(Z_{ik,j}|Z_{ik,j}^{(m)}) &=I(Y_{ik} = j) \Bigg[ Z_{ik,j}^{(m)} - \log\Big\{\sum_{j = 1}^{J_k}\exp(Z_{ik,j}^{(m)})\Big\}\Bigg]\\
     &\quad+ (Z_{ik,j} - Z_{ik,j}^{(m)})\{y_{ik,j} - \frac{\exp(Z_{ik,j}^{(m)})}{\sum_{j = 1}^{J_k}\exp(Z_{ik,j}^{(m)})}\} \\
     &\quad + \frac{1}{2}M_k(Z_{ik,j} - Z_{ik,j}^{(m)})^2,
\end{align*}
where $M_k$ is a constant bounding the curvature, and 
$\tilde{Z}_{ik,j}^{(m+1)}$ is the pseudo-response, updated as:
\begin{equation} \label{eq: update Z tilde}
     \tilde{Z}_{ik,j}^{(m+1)} = Z_{ik,j}^{(m)} - \frac{1}{M_k} \left( y_{ik,j} - p_{ik,j}^{(m)} \right).
\end{equation}
Here, $y_{ik,j} = I(Y_{ik} = j)$, $p_{ik,j}^{(m)}$ is the predicted probability for category $j$ using $Z_{ik,j}^{(m)}$. The constant $M_k$ is the derived lower bound for the smallest eigenvalue of the Hessian matrix $\nabla^2f(Z_{ik,j})$, and its value depends on the number of classes, $J_k$. 
\end{prop}

The proof of Proposition \ref{prop:mm} is given in the Appendix \ref{appendix: Update pseudo response}.
In the proof, we showed that the constant $M_k$ can take values as:  $-1/4$ if $J_k = 2$, $-1/2$ if $J_k = 3$, and $-1$ otherwise. 

We propose an iterative algorithm to optimize the penalized composite log likelihood. At each iteration $(m+1)$, given the current parameter estimate $\tilde{\boldsymbol{\theta}}^{(m)}$, the update is carried out in two main steps:

\paragraph*{1. Update the Pseudo-Response}~\\
First, we compute a pseudo-response $\tilde{Z}_{ik,j}^{(m+1)}$, based on current linear predictors $Z_{ik,j}^{(m)} = \tilde{X}_{ik,j}\tilde{\boldsymbol{\theta}}^{(m)}$. This is a gradient-based update following Equation \eqref{eq: update Z tilde}.

\paragraph*{2. Update the Parameters}~\\
Second, with this new pseudo-response, the next parameter estimate, $\tilde{\boldsymbol{\theta}}^{(m+1)}$, is obtained by solving the following penalized least squares problem:
\begin{equation*}
    \tilde{\boldsymbol{\theta}}^{(m+1)} = \underset{\tilde{\boldsymbol{\theta}}}{\text{argmin}} \left\{ \frac{1}{2} \sum_{i = 1}^n\sum_{k = 1}^K\sum_{j = 1}^{J_k} \left( \tilde{Z}_{ik,j}^{(m+1)} - \tilde{\mathbf{X}}_{ik,j} \tilde{\boldsymbol{\theta}} \right)^2 + \lambda \sum_{g=1}^G c_g \|\tilde{\boldsymbol{\theta}}_{A_g}\|_1^\gamma \right\}.
\end{equation*}
This is a standard group-penalized regression problem that can be solved efficiently. These two steps are repeated until the parameter estimates converge. The final MM algorithm is as follows:
\begin{algorithm}[h]
\caption{MM algorithm for multivariate multinomial logit}
\begin{algorithmic}[1]
\State Initialize $Z_{ik,j}$ for every $k$ and $j$
\State Set $m = 0$ and $\tilde{Z}_{ik,j}^{(0)} \gets Z_{ik,j}^{(0)}$
\Repeat
    \State Update $\tilde{Z}_{ik,j}^{(m+1)}$ for every $i,k$ and $j$ using $Z_{ik,j}^{(m)}$
    \State Update $\tilde{\boldsymbol{\theta}}^{(m+1)}$
    \State Update $Z_{ik,j}^{(m+1)}$ based on the updated $\tilde{\boldsymbol{\theta}}^{(m+1)}$
    \State $m\gets m + 1$
\Until{convergence criterion is met}
\State \Return $\tilde{\boldsymbol{\theta}}$
\end{algorithmic}
\end{algorithm}

\subsection{Tuning parameter selection} \label{sec: tuning parameter selection}

To obtain a parsimonious and predictive model, the selection of the tuning parameter ($\lambda$) is critical. We select this parameter using a modified Bayesian Information Criterion (BIC), chosen to balance model fit with model complexity.

The BIC is calculated for a range of $\lambda$ values, and we select the one that minimizes the following criterion:
\begin{equation} \label{eq: BIC}
    \text{BIC}(\lambda) = \frac{ \ell^c(\hat{\theta}_\lambda;y)}{\ell^c(\hat{\theta}_{\lambda=0};y)} + \frac{\log(n)d(\lambda)}{n}.
\end{equation}
This criterion consists of two main parts. The first term measures the goodness-of-fit, defined as the ratio of the conditional log likelihood $\ell^c(\hat{\theta}_\lambda;y)$ to that of the unpenalized model (i.e., where $\lambda$ = 0). This ratio quantifies how well the penalized model fits the data relative to the full model. Similar tuning parameter selection approaches have been used in the literature \citep[e.g.][]{lu2013variable}. In a high-dimensional setting, when $p > n$, the unpenalized estimator may not be unique. In such instances, $\ell^c(\hat{\theta}_{\lambda = 0};y)$ can be replaced by the conditional log likelihood of the saturated model with a minimal $\ell_2$ shrinkage to ensure numerical stability.
The second term is a penalty for model complexity, where $d(\lambda)$ represents the effective degrees of freedom of the model for a given $\lambda$. Following \cite{huang2009group}, the degrees of freedom term is defined as:
\begin{equation*}
    d(\lambda) = \text{tr}\left\{ \mathbf{X}_{\lambda} \left( \mathbf{X}^T_{\lambda} \mathbf{X}_{\lambda} + 0.5\mathbf{W}_{\lambda} \right)^{-1} \mathbf{X}^T_{\lambda} \right\}.
\end{equation*}
Here, $\textbf{X}_\lambda$ is the submatrix of the design matrix containing only the columns for the active predictors (i.e., those with non-zero coefficients at a penalty level $\lambda$). The term $\textbf{W}_\lambda$ is a diagonal matrix of weights used to adjust the complexity calculation, with its diagonal elements defined as:
\begin{equation*}
    \text{diag}(\textbf{W}_\lambda) =  \sum_{s \in A_g}\frac{\hat{\eta}_{p}^{1-1/\gamma}c_j^{1/\gamma}I(\hat{\theta}_{s} \ne 0)}{|\hat{\theta}_{s}|}.
\end{equation*}
where $\eta_p = c_p(\frac{1-\gamma}{\tau_n\gamma})\|\theta_{A_p}\|_1^\gamma$, and $\theta_s$ is $s$-th component of $\hat{\theta}$.
Our procedure involves evaluating the BIC score over a predefined grid of candidate $\lambda$ values. The value of $\lambda$, which has the minimum BIC score, is chosen as it represents the best trade-off between model fit and parsimony.

\section{Simulation} \label{sec: simulation}
In this section, we conducted a simulation study to evaluate the performance of our proposed method. The primary focus is on the algorithm's ability to perform simultaneous variable selection at both the group and individual levels. The BIC is used to select the optimal tuning parameter in all scenarios.

\subsection{Simulation Framework}
We establish a fixed simulation setting with $K=3$ multinomial outcomes. The number of categories for these outcomes is $J_1=2$, $J_2=3$, and $J_3=2$, respectively. The true values for the intercept parameters, $\tilde{\boldsymbol{\alpha}} = (0.8, 0.4, 1, 0.6)^T$ and $\tilde{\boldsymbol{\psi}} = (-0.9, -0.7, -1, -0.8, -0.7)^T$, are held constant across all simulations. The slope coefficients ($\tilde{\boldsymbol{\beta}}$ and $\tilde{\boldsymbol{\delta}}$), which are subject to selection, are organized by covariate into groups. To test selection performance, we designate four ``active'' groups (1, 2, 7, and 8) that contain non-zero coefficients, while all coefficients in the remaining groups are set to zero. For each of the active groups, the coefficient vector containing both $\boldsymbol{\beta}$ and $\boldsymbol{\delta}$ was specifically designed to include zero-valued elements to test for accurate individual-level variable selection.

For each simulation, the covariate matrix $\mathbf{X}$ is generated from a multivariate normal distribution with mean 0, variance 1, and covariance 0.25. The outcome vector $y_i$ is then generated by assigning it to the category combination with the highest linear predictor value. Each experimental setup is replicated 100 times to ensure stable results.

\subsection{Simulation Scenarios}
We investigate the method's performance across five distinct scenarios based on variations of the baseline coefficient matrix. The baseline coefficient is designed to assess the model's performance in variable selection at both the across-group and within-group levels. For this baseline, we specified four active groups (groups 1, 2, 7, and 8) and set the coefficient vector for all remaining groups to a zero vector, designating them as inactive. 
The true non-zero coefficient vector for the active groups were specified in the vector $(\tilde{\boldsymbol{\beta}}^T, \tilde{\boldsymbol{\delta}}^T)^T$, where elements are as follows: $(0.7, 0.8, 0.9, 0.7, 0, 0, 0, 0, 0)^T$ for Group 1; $(0.8, 1.1, 0.7, 1, -1, 0, -1, -1.2, 0)^T$ for Group 2; $(1.1, 0.7, 1.2, 0.9, -1, 0, -0.7, -1.1, 0.0)^T$ for Group 3; and $(0.8, 0.9, 0.6, 0.7, 0, -1, 0, 0, 0)^T$ for Group 8.
This simulation design allows for the assessment of the model's capacity for identifying the inactive group and identifying the true-zero-valued coefficient within the active groups.

\paragraph*{Scenario 1 (Baseline)}~\\ 
Data is generated using the baseline coefficient matrix to test both group-level and individual-level selection.

\paragraph*{Scenario 2 (Inactive Main Effect Component)} ~\\
This scenario uses the baseline coefficients but sets the second entry of the coefficient vector to zero for every active group. This second entry corresponds to the parameter representing the main effect of the covariate when the second outcome chooses the second category. 

\paragraph*{Scenario 3 (Strong Signal)} ~\\
To test performance with stronger effects, we increase the magnitude of each non-zero coefficient in the baseline coefficient matrix by 1.0.

\paragraph*{Scenario 4 (Weak Signal)}~\\
To test performance with weaker effects, we decrease the magnitude of each non-zero coefficient in the baseline coefficient matrix by 0.5.

\paragraph*{Scenario 5 (Incomplete Data)}~\\
This scenario tests the model's robustness by using the coefficients from Scenario 2 but generating data in which an outcome combination, $y_i=(2,3,2)$, is never observed. This evaluates the model's ability to estimate unique parameters when some outcome combinations are missing.

\subsection{Simulation Result}
To evaluate our method's performance, we assessed both estimation accuracy and variable selection capability. Estimation accuracy was measured by the Mean Squared Error (MSE), calculated as the average squared Euclidean distance between the estimated ($\hat{\boldsymbol{\theta}}$) and true ($\boldsymbol{\theta}$) parameter vectors over 100 replications:
\begin{equation*}
    \text{MSE} = \frac{1}{100}\sum_{s = 1}^{100}\|\hat{\boldsymbol{\theta}}^{(s)} - \boldsymbol{\theta}\|_2^2.
\end{equation*}
The model's ability to correctly identify individual coefficients was quantified by the True Positive Rate (TPR) and True Negative Rate (TNR), defined as:
\begin{align*}
    \text{TPR} &= \frac{\sum_{i} I(\theta_{i} \neq 0 \text{ and } \hat{\theta}_i \neq 0)}{\sum_{i} \mathbb{I}(\theta_{i} \neq 0)}, \\
    \text{TNR} &= \frac{\sum_{i} I(\theta_{i} = 0 \text{ and } \hat{\theta}_i = 0)}{\sum_{i} \mathbb{I}(\theta_{i} = 0)}.
\end{align*}
We also evaluated selection at the group level by counting the correctly identified active groups ($\text{TP}_{\text{g}}$) and inactive groups ($\text{TN}_{\text{g}}$), where $\|\theta_{A_p}\|_0$ is the number of non-zero elements in a group:
\begin{align*}
    \text{TP}_{\text{g}} &= \sum_{g=1}^G I(\|\theta_{A_g}\|_0 > 0 \text{ and } \|\hat{\theta}_{A_g}\|_0 > 0), \\
    \text{TN}_{\text{g}} &= \sum_{g=1}^G I(\|\theta_{A_g}\|_0 = 0 \text{ and } \|\hat{\theta}_{A_g}\|_0 = 0).
\end{align*}
Lastly, we assess the variable selection performance specifically within the true active groups. We define the set of active groups as $\mathcal{A} = \{g\;;\;\|\boldsymbol{\theta}_{A_g}\|_0 > 0\}$. Within this set, we count the number of correctly identified true non-zero coefficients ($\text{TP}_{\text{wi.g}}$) and true zero coefficients ($\text{TP}_{\text{wi.g}}$) in the true active group as follows:
\begin{align*}
\text{TP}_{\text{wi.g}} &= \sum_{g \in \mathcal{A}} \sum_{i \in A_g} I(\theta_{i} \neq 0 \text{ and } \hat{\theta}_i \neq 0), \\
\text{TN}_{\text{wi.g}} &= \sum_{g \in \mathcal{A}} \sum_{i \in A_g} I(\theta_{i} = 0 \text{ and } \hat{\theta}_i = 0).
\end{align*}

\begin{table}[h!] 
    \centering
    \caption{Simulation results showing the mean (standard deviation) calculated over 100 runs. For reference, the ideal value for group-level true negatives ($\text{TN}_g$) is 6 (for $p=10$) and 16 (for $p=20$), while the total number of active groups ($\text{TP}_g$) is 4 in all scenarios. The maximum possible scores for ($\text{TN}_{\text{wi.g}}$, $\text{TP}_{\text{wi.g}}$) are (13, 23) for scenarios 1, 3, and 4, and (17, 19) for scenarios 2 and 5, respectively.}

    \renewcommand{\arraystretch}{1.7}

    \resizebox{\textwidth}{!}{
    \begin{tabular}{c c c c c c c c c c}
    \hline
Scenario & n & p & MSE & TNR & TPR & $\text{TN}_g$& $\text{TP}_g$ & $\text{TN}_{wi.g}$ & $\text{TP}_{wi.g}$\\
\hline
\multirow{ 6}{*}{Scenario 1} & \multirow{ 2}{*}{500} & 10 & 25.061(7.47) & 0.965(0.02) & 0.551(0.15) & 6(0) & 2.46(0.82) & 10.67(1.58) & 8.65(4.93)\\
 && 20 & 20.463(9.26) & 0.979(0.01) & 0.656(0.19) & 15.99(0.1) & 2.98(0.89) & 9.72(1.99) & 11.98(5.98)\\
 \cline{2-10}
&\multirow{ 2}{*}{1000}  & 10 & 4.539(3.51) & 0.906(0.03) & 0.976(0.06) & 6(0) & 3.99(0.1) & 6.67(1.88) & 22.24(1.77)\\
&& 20 & 3.872(1.88) & 0.959(0.01) & 0.985(0.02) & 16(0) & 4(0) & 6.51(1.79) & 22.51(0.66)\\
\cline{2-10}
&\multirow{ 2}{*}{2000}  & 10 & 1.759(1.25) & 0.897(0.03) & 0.998(0.01) & 5.99(0.1) & 4(0) & 6.15(1.89) & 22.95(0.26)\\
& & 20 & 1.826(1) & 0.956(0.01) & 0.998(0.01) & 16(0) & 4(0) & 6.08(1.87) & 22.94(0.28)\\
\hline
\multirow{ 6}{*}{Scenario 2} & \multirow{ 2}{*}{500} & 10 & 13.79(3.94) & 0.981(0.02) & 0.58(0.16) & 6(0) & 2.06(0.96) & 15.68(1.63) & 7.23(4.51)\\
 & & 20 & 11.142(4.73) & 0.984(0.01) & 0.691(0.18) & 16(0) & 2.69(0.94) & 14.38(2.11) & 10.34(5.13)\\
\cline{2-10}
& \multirow{ 2}{*}{1000}  & 10 &4.028(2.89) & 0.927(0.03) & 0.942(0.09) & 6(0) & 3.8(0.4) & 11.82(2.24) & 17.38(2.55)\\
& & 20 & 3.553(2.74) & 0.966(0.01) & 0.951(0.09) & 15.99(0.1) & 3.83(0.38) & 11.47(2.27) & 17.62(2.46)\\
\cline{2-10}
& \multirow{ 2}{*}{2000}  & 10 & 1.388(0.67) & 0.915(0.03) & 0.999(0.01) & 6(0) & 4(0) & 10.99(2.17) & 18.98(0.14)\\
 & & 20 & 1.317(0.58) & 0.961(0.01) & 1(0) & 16(0) & 4(0) & 10.66(1.71) & 18.99(0.1)\\
\hline
\multirow{ 6}{*}{Scenario 3} & \multirow{ 2}{*}{500} & 10 & 17.118(9.16) & 0.87(0.03) & 0.995(0.01) & 5.92(0.27) & 4(0) & 4.41(1.79) & 22.85(0.39)\\
 & & 20 & 19.233(8.97) & 0.943(0.01) & 0.996(0.01) & 15.81(0.42) & 4(0) & 4.59(1.82) & 22.88(0.33)\\
\cline{2-10}
& \multirow{ 2}{*}{1000} & 10 & 8.346(3.52) & 0.866(0.02) & 1(0) & 5.96(0.2) & 4(0) & 4.05(1.55) & 23(0)\\
& & 20 & 8.269(4.47) & 0.944(0.01) & 1(0) & 15.94(0.24) & 4(0) & 4.31(1.58) & 23(0)\\
\cline{2-10}
& \multirow{ 2}{*}{2000} & 10 & 3.99(2.12) & 0.864(0.03) & 1(0) & 5.99(0.1) & 4(0) & 3.96(1.63) & 23(0)\\
& & 20 & 4.25(2.22) & 0.943(0.01) & 1(0) & 15.96(0.2) & 4(0) & 4.11(1.66) & 23(0)\\
\hline
\multirow{ 6}{*}{Scenario 4} & \multirow{ 2}{*}{500} & 10 & 5.873(0.93) & 0.998(0.01) & 0.293(0.03) & 6(0) & 0.17(0.38) & 12.86(0.38) & 0.39(1.13)\\
 & & 20 & 5.821(1.15) & 0.998(0) & 0.309(0.05) & 16(0) & 0.32(0.51) & 12.69(0.58) & 0.9(1.59)\\
\cline{2-10}
& \multirow{ 2}{*}{1000} & 10 & 4.907(0.95) & 0.986(0.01) & 0.382(0.06) & 6(0) & 0.89(0.42) & 12.06(0.75) & 3.23(2)\\
& & 20 & 4.709(0.83) & 0.993(0) & 0.389(0.05) & 16(0) & 0.95(0.3) & 11.84(0.75) & 3.44(1.61)\\
\cline{2-10}
& \multirow{ 2}{*}{2000} & 10 & 3.875(0.76) & 0.978(0.01) & 0.469(0.08) & 6(0) & 1.26(0.48) & 11.55(0.73) & 6.01(2.39)\\
&  & 20 & 3.656(0.86) & 0.991(0.01) & 0.487(0.09) & 16(0) & 1.32(0.58) & 11.53(0.8) & 6.58(3.02)\\
\hline
\multirow{ 6}{*}{Scenario 5} & \multirow{ 2}{*}{500} & 10 & 10.645(3.9) & 0.969(0.02) & 0.717(0.16) & 6(0) & 2.71(0.88) & 14.78(1.73) & 11.07(4.41)\\
 & & 20 & 9.884(4.28) & 0.985(0.01) & 0.749(0.17) & 16(0) & 2.93(0.87) & 14.58(1.82) & 11.96(4.79)\\
\cline{2-10}
& \multirow{ 2}{*}{1000} & 10 & 3.702(1.55) & 0.936(0.03) & 0.974(0.05) & 6(0) & 3.94(0.24) & 12.45(1.97) & 18.28(1.50)\\
& & 20 & 3.812(1.86) & 0.971(0.01) & 0.975(0.06) & 16(0) & 3.93(0.26) & 12.4(1.99) & 18.3(1.68)\\
\cline{2-10}
& \multirow{ 2}{*}{2000} & 10 & 2.529(0.51) & 0.927(0.03) & 0.999(0.01) & 6(0) & 4(0) & 11.83(2.09) & 18.97(0.17)\\
 & & 20 & 2.48(0.52) & 0.967(0.01) & 0.999(0.01) & 16(0) & 4(0) & 11.61(2.08) & 18.98(0.14)\\
\hline
    \end{tabular}
    }
    \label{tab: simulation result}
\end{table}

The simulation results demonstrate the robust performance and convergence of the proposed algorithm in Table \ref{tab: simulation result}. A key finding across all scenarios is that the MSE consistently decreased as the sample size increased, indicating that model accuracy improves with more data. This trend was mirrored in the variable selection performance, where both the TPR and TNR approached or exceeded 0.9 with larger sample sizes. The notable exception was the weak signal scenario, where the low signal strength hindered the identification of true positives, resulting in a lower TPR. Nevertheless, even in this challenging scenario, the model maintained excellent specificity, with a TNR consistently above 0.9. At the group level, the algorithm demonstrated near-perfect performance in identifying inactive groups ($\text{TN}_g$) across all settings, and active group selection ($\text{TP}_g$) was also highly effective, except again in the weak signal scenario. Within the correctly identified active groups, the model's ability to pinpoint true non-zero coefficients ($\text{TP}_{\text{wi.g}}$) improved with sample size. Interestingly, in the strong signal scenario, the model achieved perfect identification of true positive elements but also tended to include some zero-valued coefficients as false positives. For the incomplete data, the MSE also decreases as $n$ increases, and the proposed model effectively selects the variable across groups and within the group, as it provides a unique parameter estimation. Overall, the proposed algorithm effectively selects non-zero coefficients at both the group and within-group levels and is notably robust in scenarios with missing data combinations.

In addition to evaluating the absolute performance of the proposed model, previously assessed via MSE, and variable selection performance, we designed a comparative study to further validate our approach. To establish a baseline, we utilized a marginal model that assumes independence among outcomes. To ensure a fair comparison, a group bridge penalty was also applied to this baseline model. In the marginal model, the linear predictor, $Z_{ik,j}'$, is formulated as
\begin{equation*}
    Z_{ik,j}' = \alpha_{k,j} + X_i\beta_{k, j},
\end{equation*}
yielding the corresponding conditional probability:
\begin{equation*}
    P(Y_{ik} = j|X_i) = \frac{\text{exp}(Z_{ik,j}')}{\sum_{h = 1}^{J_k}\text{exp}(Z_{ik,h}')}.
\end{equation*}
For both models, the tuning parameters were selected using the BIC, as defined in Equation \eqref{eq: BIC}. We restricted our comparative evaluation specifically to Scenario 5 (Incomplete Data). 

The primary performance evaluation criterion was the Area Under the Curve (AUC) for multinomial outcomes, defined as:
\begin{equation} \label{eq: AUC}
    \text{AUC}= \frac{1}{n(K-1)}\sum_{k = 1}^K\sum_{\{i:y_{ik} = j\}}\Bigg\{\sum_{s \ne j}I(P_{ik,j} > P_{ik,s})\Bigg\},
\end{equation}
where $P_{ik,j}$ is the predicted probability that outcome $k$ for individual $i$ is category $j$. This AUC metric quantifies how often the model correctly assigns the highest probability to the true outcome category. We evaluated two distinct types of AUC metrics: conditional AUCs and a joint AUC. The conditional AUCs were calculated to measure the predictive accuracy for each individual outcome ($y_1,y_2$ and $y_3$). Additionally, to evaluate the overall multivariate predictive performance, we computed a joint AUC by combining the three outcomes into a unique combination representation. These metrics, along with their standard deviations, were compared over 100 simulation replications. For each replication, the AUC values were computed using an independent test set comprising 10,000 observations, generated under the identical data generation process as the training data. 

\begin{table}[h!] 
    \centering 
    
    \caption{Simulation results under Scenario 5 (Incomplete data setting): Comparison of average AUC values and standard deviations between the proposed model (mml) and the marginal model. Results are presented across different sample sizes ($n \in \{500, 1000, 2000\}$) and covariate dimension ($p \in \{10, 20\}$) based on 100 replications. $y_1,y_2$ and $y_3$ denote the conditional AUC for each outcome, and `joint' represents the joint AUC calculated using the unique combination representation of the three outcomes.}

    \renewcommand{\arraystretch}{1.2}
    
    \resizebox{0.9\textwidth}{!}{
    \begin{tabular}{c  c  c  c  c c c}
    \hline
    n&p & model & $y_1$ & $y_2$ & $y_3$ & joint\\
    \hline
    \multirow{ 4 }{*}{ 500 } & \multirow{ 2 }{*}{ 10 } & mml & 0.674(0.032) & 0.693(0.029) & 0.7(0.033) & 0.711(0.029)\\
    & & marginal & 0.641(0.022) & 0.677(0.015) & 0.665(0.013) & 0.688(0.017)\\
    \cline{2-7}
    & \multirow{ 2 }{*}{ 20 }  & mml & 0.674(0.033) & 0.697(0.025) & 0.697(0.029) & 0.712(0.025)\\
    & & marginal & 0.64(0.024) & 0.68(0.013) & 0.661(0.014) & 0.689(0.018)\\
    \hline
    \multirow{ 4 }{*}{ 1000 } & \multirow{ 2 }{*}{ 10 } & mml &0.718(0.009) & 0.721(0.006) & 0.743(0.008) & 0.744(0.005)\\
    & & marginal & 0.657(0.007) & 0.687(0.003) & 0.673(0.004) & 0.703(0.003)\\
    \cline{2-7}
    & \multirow{ 2 }{*}{ 20 } & mml & 0.715(0.01) & 0.724(0.006) & 0.732(0.01) & 0.741(0.006)\\
    & & marginal & 0.661(0.007) & 0.688(0.004) & 0.668(0.004) & 0.703(0.004)\\
    \hline
    \multirow{ 4 }{*}{ 2000 } & \multirow{ 2 }{*}{ 10 } & mml  & 0.724(0.003) & 0.726(0.004) & 0.748(0.003) & 0.749(0.002)\\
    & & marginal & 0.659(0.003) & 0.689(0.002) & 0.674(0.004) & 0.707(0.003)\\
    \cline{2-7}
    & \multirow{ 2 }{*}{ 20 } & mml & 0.721(0.002) & 0.729(0.004) & 0.739(0.003) & 0.746(0.002)\\
    & & marginal & 0.663(0.005) & 0.691(0.004) & 0.669(0.003) & 0.707(0.003)\\
    \hline
    \end{tabular}
    }
    \label{tab: Simulation prediction performance}
\end{table}

As illustrated in Table \ref{tab: Simulation prediction performance}, the simulation results revealed a clear performance trajectory associated with the sample size for both conditional and joint AUC. As the sample size increased from $n = 500$ to $n = 2000$, both models exhibited an increase in overall average AUC values coupled with a notable decrease in variance. Furthermore, the proposed model consistently outperformed the comparison model, yielding higher AUC values across the evaluations. This confirms that the proposed method maintains superior predictive capability and robustness, particularly under the challenging conditions of incomplete data scenarios.

\section{Real Data Analysis} \label{sec: real data analysis}

Using data from the AURORA study, we now apply our proposed algorithm to identify key risk factors associated with correlated multiple multinomial posttraumatic disorders. The Aurora study is a large-scale national research initiative designed to understand and predict health disorders following trauma exposure \citep{mclean2020aurora}. In the study, trauma survivors, most of whom experienced a car accident, completed weekly surveys to report on their recovery process.

While the original study includes many posttraumatic disorders, our analysis focuses on three highly co-occurring disorders: Pain, Depression, and Somatic symptoms. For each disorder, using the growth mixture model (mixture linear mixed model), subgroups of trauma survivors were identified based on their trajectory pattern within the first 8 weeks after trauma exposure \citep{beaudoin2023use}. These subgroups represent subtypes of trauma survivors with different recovery processes and will be used as the outcomes for this study. To ensure sufficient sample sizes for analysis, some of the original subgroups were merged. The final subgroups for the three outcomes are as follows:
\begin{align*}
    &Y_{\text{Pain}} \in \{
        \text{Low},\;
        \text{Recovery},\;
        \text{Middle},\;
        \text{High}\}, \\
    &Y_{\text{Depression}} \in
\{        \text{Low},\;
        \text{Middle/Worsening}\},\\
    &Y_{\text{Somatic}} \in \{
       \text{Low/Recovery},
        \text{Middle/Worsening} \}.
\end{align*}

\begin{figure}[h!]
    \centering
    
    \caption{The alluvial plot illustrates the flow of participants between their recovery patterns in three different symptoms. The width of the ribbons represents the number of participants.}\includegraphics[width=0.8\linewidth]{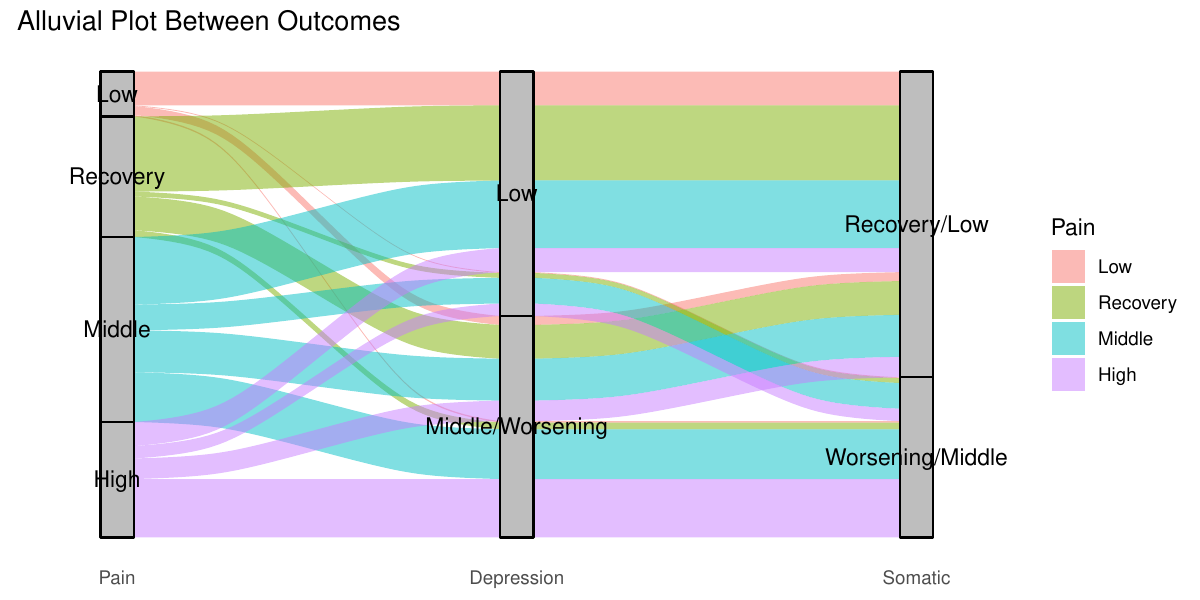}
   
    \label{fig:alluvial}
\end{figure}

An alluvial plot was utilized to illustrate the correspondence between the categories of our multinomial outcomes (Figure \ref{fig:alluvial}). The plot shows that participants who reported a 'Low' pain level were more likely to also report the lowest severity categories for Depression and Somatic symptoms. This visualization confirms that the three outcomes are highly correlated.

\begin{table}[h!] 
    \centering 
    
    \caption{Parameter estimates from the proposed model applied to the Aurora study. Each row details the estimated coefficient ('Value') for a covariate, conditional on specific levels of the three multinomial outcomes.}

    \renewcommand{\arraystretch}{1.2}
    
    \resizebox{0.9\textwidth}{!}{
    \begin{tabular}{c c c c c c}
    \hline
        Coefficient &$Y_{\text{Pain}}$ & $Y_{\text{Depression}}$ & $Y_{\text{Somatic}}$ & Covariate & Value  \\
        \hline
        \multirow{ 5 }{*}{ $\boldsymbol{\alpha}$ }& Recovery& & & & 1.092\\
        & Middle & & & & 0.714\\
        & High & & & & -1.068\\
        & & Middle/Worsening& & & -4.455\\
        & & & Middle/Worsening & & -3.405\\
        \hline
        \multirow{ 7 }{*}{$\boldsymbol{\psi}$}& Recovery & Middle/Worsening &  & & -0.101\\
        & Middle & Middle/Worsening & & & 0.814\\
        & High & Middle/Worsening & & & 1.549\\
        & Recovery & & Middle/Worsening & & 0.642\\
        & Middle & & Middle/Worsening & & 2.739\\
        & High & & Middle/Worsening & & 2.900\\
        & & Middle/Worsening & Middle/Worsening & & 1.491\\
        \hline
        \multirow{ 13 }{*}{$\boldsymbol{\beta}$}& Recovery & & & Age & -0.008 \\
        \cline{2-6}
        & Middle & & & Pain & 0.006 \\
        \cline{2-6}
        & High & & & Age & 0.004 \\
        & High & & & Pain & 0.136 \\
        & High & & & Black & 0.135 \\
        \cline{2-6}
        & & Middle/Worsening & & Age & -0.010 \\
        & & Middle/Worsening & & Depression & 0.069 \\
        & & Middle/Worsening & & Insomnia & 0.036 \\
        & & Middle/Worsening & & Pain & -0.013 \\
        & & Middle/Worsening & & Black & -0.251 \\
        \cline{2-6}
        & & & Middle/Worsening & Depression & 0.000 \\
        & & & Middle/Worsening & Insomnia & -0.008 \\
        & & & Middle/Worsening & Pain & 0.020 \\
        \hline
        \multirow{ 13}{*}{$\boldsymbol{\delta}$}& Recovery & Middle/Worsening & & Depression & 0.011 \\
        & Middle & Middle/Worsening & & Pain & 0.008 \\
        & Middle & Middle/Worsening & & Black & -0.110 \\
        & High & Middle/Worsening & & Pain & -0.047 \\
        & High & Middle/Worsening & & Black & 0.002 \\
        \cline{2-6}
        & Recovery & & Middle/Worsening & Pain & 0.002 \\
        & Middle & & Middle/Worsening & Age & -0.019 \\
        & Middle & & Middle/Worsening & Pain & 0.007 \\
        & Middle & & Middle/Worsening & Black & 0.401 \\
        \cline{2-6}
        & & Middle/Worsening & Middle/Worsening & Age & 0.000 \\
        & & Middle/Worsening & Middle/Worsening & Depression & -0.005 \\
        & & Middle/Worsening & Middle/Worsening & Insomnia & -0.001 \\
        & & Middle/Worsening & Middle/Worsening & Black & 0.246 \\
        \hline
    \end{tabular}
    }
    \label{tab:Variable Selection}
\end{table}

Applying the proposed algorithm resulted in the identification of several significant predictors. We used 5-fold cross-validation for tuning parameter selection, as we found that the BIC produced an overly sparse model that obscured the relationships between outcomes. The estimated association parameters ($\psi$) were generally positive, confirming the strong positive correlations observed in the alluvial plot. For instance, a positive $\psi$ implies that, for a given set of covariates, the two corresponding outcome categories are more likely to be chosen together.

The model identified Age, Black race, pre-trauma Pain, Depression, and Insomnia as significant predictors. For the estimated slope coefficients ($\boldsymbol{\beta}$ and $\boldsymbol{\delta}$), a positive value indicates that a one-unit increase in the corresponding covariate increases the log-odds ratio, making the associated outcome categories more likely to be selected together. A negative value implies the opposite.

To quantify the predictive performance of the proposed algorithm, we conducted a rigorous cross-validation study on the Aurora Dataset. For a robust evaluation, the entire process was repeated 10 times. In each replication, the data were first split into a training and a test set. The training data was then further split into 5 folds to select the optimal tuning parameter, $\lambda$, via cross-validation. The value of $\lambda$ that maximized the composite likelihood on the validation folds was chosen as optimal. The model was then refit on the entire training dataset using this optimal $\lambda$, and its final predictive performance was evaluated on the test data.

\begin{figure}[h!] 
    \centering

    \caption{These boxplots summarize the distribution of Area Under the Curve (AUC) values from 10 replications, comparing the predictive accuracy of our proposed method (mml) with a standard marginal model. The three panels on the left show the conditional AUCs for predicting each of the three individual outcomes: Pain, Depression, and Somatic Symptoms. The panel on the right shows the AUC for predicting the joint probability of all three outcomes combined. In all cases, a higher AUC indicates better model performance.}
    \includegraphics[width=0.7\linewidth]{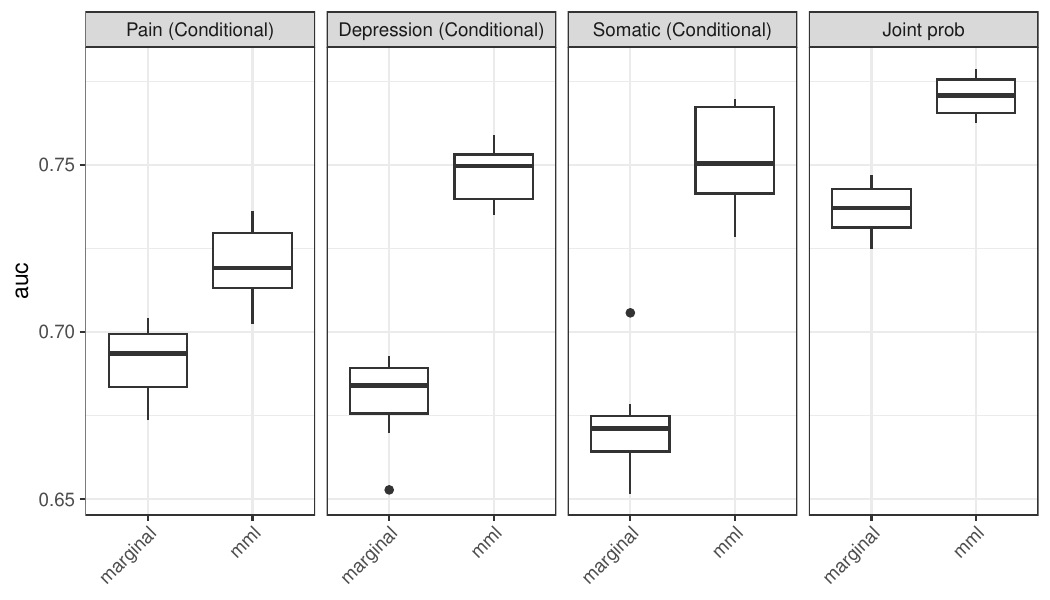}
    
    \label{fig: AUC boxplot}
\end{figure}

The predictive performance for the multinomial outcome was primarily quantified via the AUC as defined in Equation \ref{eq: AUC}. We first compared the predictive performance based on conditional probabilities. The comparison model was a marginal model that fit a separate multinomial logistic regression to each outcome independently. Specifically, this marginal model ignored the interdependence between outcomes by omitting the association parameters, while applying the same group bridge penalty. As shown in Figure \ref{fig: AUC boxplot}, the proposed algorithm achieved a higher AUC than the marginal model for all three outcomes, indicating superior predictive performance.

We also evaluated the model's ability to predict the joint distribution of the outcomes. Because the composite likelihood yields a consistent estimator for the joint likelihood, its predictive accuracy for the joint outcome is of key interest. For a baseline comparison, we implemented a marginal model by fitting a standard multinomial logistic regression to a single combined response variable. This new variable had $4 \times 2 \times 2=16$ distinct levels, each representing a unique combination of the three outcomes. As shown in Figure \ref{fig: AUC boxplot}, our proposed algorithm again produced higher AUC values than the marginal fit. This result supports the consistency of our composite likelihood approach, demonstrating its effectiveness in predicting the joint distribution.

\section{Conclusion}\label{sec:discussion}

In this paper, we have introduced a multivariate multinomial logit model that uses ANOVA decomposition to capture outcome correlations while reducing the dimension of the parameter space. This model offers several key advantages: it maintains a parsimonious parameter space, its parameters have a clear interpretation, and it is computationally feasible to estimate. For parameter estimation, we proposed an efficient procedure that combines a composite likelihood function to manage the computational burden, a group bridge penalty for variable selection, and an MM algorithm for optimization. This framework yields consistent estimators and identifies significant predictors.

The performance of our method was evaluated through extensive simulations and a real-data application. A Monte Carlo study confirmed that the proposed algorithm effectively selects non-zero coefficients at both the group and individual levels and is notably robust, even in scenarios with missing data combinations. Furthermore, the algorithm was applied to the Aurora study, a dataset featuring highly correlated clinical outcomes. The model identified Age, Black race, and baseline levels of Pain, Depression, and Insomnia as significant predictors. A cross-validation analysis demonstrated that our approach achieves superior predictive accuracy, measured by the AUC, when compared to fitting separate marginal models for each outcome.

Finally, the proposed model can be extended in several directions. A promising avenue for future research is the incorporation of functional covariates. The increasing availability of dense, longitudinal data from sources such as electronic health records and wearable devices presents an opportunity to enhance the model's predictive capabilities. Expanding the framework to leverage this type of complex data could significantly improve its clinical utility.

\bibliographystyle{apalike}
\bibliography{reference}

\appendix

\section{Connection between conditional and joint probability} \label{appendix: connection between conditional and joint probability}

This appendix elaborates on the connection between conditional and joint probability, as discussed in Section \ref{sec: multivariate multinomial model} of the main paper. We will derive the joint probability from the conditional model using the ANOVA decomposition of model parameters. Following \cite{besag1974spatial}, we will use the identity equation to derive the joint probability.
\begin{align} \label{eq: besag}
     \frac{P(y_1,\dots, y_K)}{P(1, \dots, 1)} 
     &= \prod_{k = 1}^K\frac{P(y_k|1, \dots, 1, y_{k+1}, \dots, y_K)}{P(1|1, \dots, 1, y_{k+1}, \dots, y_K)}.
\end{align} 
Then, the denominator in the conditional probability on the right-hand side in Equation \eqref{eq: besag} is the same. Also, the numerator of $P(1|1, \dots, 1, y_{k+1}, \dots, y_K)$ is proportional to 1 due to the identification restriction. Hence, by Equation (5) in the main paper
\begin{align}\label{eq: besag anova}
    \frac{P(y_1,\dots, y_K)}{P(1,\dots,1)} &= \prod_{k = 1}^K\text{exp}\Big\{\alpha_{k,y_k} + X\beta_{k,y_{k}}+ \sum_{\ell < k}(\psi_{k\ell, y_k1} + X\delta_{k\ell,y_k1}) \\
&\quad \quad\quad\quad+ \sum_{k < \ell}(\psi_{k\ell,y_ky_\ell }+X\delta_{k\ell,y_ky_\ell})\Big\}. \nonumber
\end{align}
Since $\psi_{k\ell, y_k1} = 0$ and $\delta_{k\ell, y_k1} = 0$, Equation \eqref{eq: besag anova} can be rewritten as 
\begin{align} \label{eq: besag anova final}
        \frac{P(y_1,\dots, y_K)}{P(1,\dots,1)} &= \prod_{k = 1}^K\text{exp}\Big\{\alpha_{k,y_k} + X\beta_{k,y_{k}} 
        + \sum_{k < \ell}(\psi_{k\ell,y_ky_\ell }+X\delta_{k\ell,y_ky_\ell})\Big\}.
\end{align}
Also, to obtain $P(y_1, \dots, y_K)$, we will use the identity
\begin{equation} \label{eq: appendix joint prob}
    P(y_1, \dots, y_K) = \frac{P(y_1, \dots, y_K)/P(1, \dots, 1)}{\sum_{s\in S}P(s_1, \dots, s_K)/P(1, \dots, 1)},
\end{equation}
where $S$ is the set of all possible choice combinations, so $\sum_{s\in S}P(s_1, \dots, s_K) = 1.$ Then, by Equation \eqref{eq: besag anova final} and \eqref{eq: appendix joint prob}, 
\begin{equation*}
    P(y_1, \dots, y_K) = \frac{\text{exp}(\mu_y)}{\sum_{s \in S}\text{exp}(\mu_s)},
\end{equation*}
where
\begin{equation*}
    \mu_y = \sum_{k = 1}^K\Big\{\alpha_{k,y_k} + X\beta_{k,y_{k}} 
        + \sum_{k < \ell}(\psi_{k\ell,y_ky_\ell }+X\delta_{k\ell,y_ky_\ell})\Big\}.
\end{equation*}

Next, we'll do the reverse, showing how to determine the conditional model from the joint probability. For simplicity, the conditional probability of $Y_{ik} = j$ given other outcomes is defined as
\begin{equation} \label{eq: appendix cond prob}
    P(y_k = j|y_{-k}) = \frac{P(y_1, \dots, y_{k-1}, y_k = j, y_{k+1}, \dots, y_K)}{P(y_{-k})},
\end{equation}
where $y_{-k} = (y_1, \dots, y_{k-1}, y_{k+1}, \dots, y_K)$. Then the denominator can be rewritten by marginalization as follows:
\begin{equation*}
    P(y_{-k}) = \sum_{s = 1}^{J_k}P(y_1, \dots, y_{k-1}, y_k = s, y_{k+1}, \dots, y_K).
\end{equation*}
Let $y^{(s)} = (y_1, \dots, y_{k-1}, y_k = s, y_{k+1}, \dots, y_K)$ for simplicity. Then, Equation \eqref{eq: appendix cond prob} can be rewritten as
\begin{equation} \label{eq: appendix cond prob using mu}
    P(y_k = j|y_{-k})= \frac{\text{exp}(\mu_{y^{(j)}})}{\sum_{s\in \{1, \dots, J_k\}}\text{exp}(\mu_{y^{(s)}})}.
\end{equation}
The denominator of the joint probability in Equation (6) in the main paper is the same in both the numerator and denominator of Equation \eqref{eq: appendix cond prob using mu}, so it is dropped out in the ratio. 
The sum of the main effects and pairwise interaction terms $\mu_y$ can be decomposed into a part that depends on $y_k$ and a part that is independent of it. 
\begin{align} \label{eq: appendix mu}
    \mu_y &= \sum_{k = 1}^K\Big\{\alpha_{k,y_k} + X\beta_{k,y_{k}} 
        + \sum_{k < \ell}(\psi_{k\ell,y_ky_\ell }+X\delta_{k\ell,y_ky_\ell})\Big\}\\
        &= \alpha_{k,y_k} + X\beta_{k,y_k} + \sum_{k < \ell}(\psi_{k\ell, y_ky_\ell} + X\delta_{k\ell, y_ky_\ell})  \\
        &\quad+\sum_{k > \ell}(\psi_{\ell k,y_\ell y_k} + X\delta_{\ell k, y_\ell y_k}) + R(-k),\nonumber
\end{align}
where $R(-k)$ is the part which does not contains $y_k$. Then, by Equation \eqref{eq: appendix cond prob using mu} and \eqref{eq: appendix mu}, the conditional probability can be rewritten as
\begin{align*}
    P(y_k = j|y_{-k}) &= \frac{\text{exp}\{Z_{ik,j} + R(-k)\}}{\sum_{s\in\{1, \dots, J_k\}}\text{exp}\{Z_{ik,s} + R(-k)\}}\\
    &= \frac{\text{exp}\{R(-k)\}\times\text{exp}\{Z_{ik,j}\}}{\text{exp}\{R(-k)\}\times\sum_{s\in\{1, \dots, J_k\}}\text{exp}\{Z_{ik,s}\}}\\
    &= \frac{\text{exp}(Z_{ik,j})}{\sum_{s = 1}^{J_k}\text{exp}(Z_{ik,s})},
\end{align*}
where 
\begin{equation*}
    Z_{ik,j} = \alpha_{k,j} + X\beta_{k,j} + \sum_{k \ne \ell}(\psi_{k\ell, jy_\ell} + X\delta_{k\ell, jy_\ell}).
\end{equation*}

\section{Proof of the Asymptotic Properties} \label{appendix: Proof of asymptotic properties}

In this section, we give the detailed proof of Theorem \ref{thm: theta distance} and Theorem \ref{thm: consistency} presented in Section \ref{subsec: asymptotic property}.

\paragraph*{Proof of Theorem \ref{thm: theta distance}} ~\\

Let $B(C) = \{\theta:\theta = \theta_0 + C\alpha_nu, \|u\|_2 = 1\}$. It is sufficient to prove that, for every $\epsilon > 0$, there exist $B(C)$ such that
\begin{equation*}
    Pr(\text{inf}_{\theta\in B(C)}Q(\theta) > Q(\theta_0)) > 1-\epsilon.
\end{equation*}
Since 
\begin{equation*}
    \frac{1}{n}Q(\theta) - \frac{1}{n}Q(\theta_0) = \frac{1}{n}[L(\theta) - L(\theta_0)] + D_n(C),
\end{equation*}
where $D_n(C) = (\lambda_n/n)(\sum_{g = 1}^Gc_g\|\theta_{A_g}\|_1^\gamma - \sum_{g = 1}^Gc_g\|\theta_{0A_g}\|_1^\gamma)$ for $\theta \in B(C)$. Then, by the Taylor expansion, we have
\begin{align*}
    \frac{1}{n}[L(\theta) - L(\theta_0)] &= \frac{1}{n}(\theta - \theta_0)'\nabla L(\theta_0) + \frac{1}{2n}(\theta - \theta_0)'\nabla^2L(\theta^*)(\theta - \theta_0) \\
    &= B_1 + B_2,
\end{align*}
where $\theta^* \in [\theta,\theta_0]$.
By the arguments in \cite{cai2005variable}, $\|\nabla L(\theta_0)\|_2 = O_p(\sqrt{nd_n})$ and $n^{-1}\nabla^2L(\theta^*) = H(\theta_0) + o_p(1)$, where $H(\theta_0) = E(n^{-1}\nabla^2L(\theta_0))$. It follows that $B_1 = O_p(C\alpha_n^2)$ and $B_2 = O_p(C^2\alpha_n^2)$ by the assumption that $H(\theta_0)$ is positive definite. If we choose sufficiently large $C$, $B_2$ dominates $B_1$ uniformly in $\|u\|_2 = 1$. For the lower bound of $D_n(C)$, it is sufficient to consider the case where $\|\theta_{0A_g}\|_1 \ge\|\theta_{Ag}\|_1$. Since $b^{\gamma} - a^\gamma \le2(b-a)b^{\gamma - 2}$ for $0 \le a \le b$ and Cauchy Schwartz,
\begin{align*}
    \sum_{j = 1}^Jc_j\Big(\|{\theta}_{0A_j}\|_1^\gamma - \|\hat{{\theta}}_{nA_j}\|_1^\gamma\Big)
&\le2\sum_{j = 1}^{J_1}c_j\|{\theta}_{0A_j}\|_1^{\gamma-1}\Big(|A_j|\|\hat{{\theta}}_{nA_j} - {\theta}_{0A_j}\|_2^2\Big)^{1/2}\\
&\le2\eta_n\Big(\sum_{j = 1}^{J_1}\|\hat{{\theta}}_{nA_j}-{\theta}_{0
A_j}\|_2^2\Big)^{1/2}\\
&\le2\eta_n\sqrt{C_n^*}\|\hat{{\theta}} - {\theta}_0\|_2,
\end{align*}
where $\eta_n = \Big(\sum_{j = 1}^{J_1}c_j^2\|{\theta}_{0A_g}\|_1^{2\gamma - 2}|A_g|\Big)^{1/2}$.
Therefore,
\begin{align} \label{eq: lower_bound of Q}
\frac{1}{n}Q(\theta) - \frac{1}{n}Q(\theta_0) 
&\ge\frac{1}{2}C^2\alpha_n^2u'[H(\theta_0) + o_p(1)]u + O_p(C\alpha_n^2) \nonumber\\
&\quad- 2\frac{\lambda_n}{n}\eta_n\sqrt{C_n^*}C\alpha_n
\end{align}
Since $-2(\lambda_n/n)\eta_n\sqrt{C_n^*}C\alpha_n$ is of order $C\alpha_n^2$, the first term of the right side of the Equation \eqref{eq: lower_bound of Q} dominates the third term uniformly in $\|u\|_2 = 1$, when $C$ is large enough. It implies that there exists a local minimizer $\hat{\theta}$ within the ball $\{\theta: \|\theta - \theta_0\| \le C\alpha_n\}$ with probability 1.

\paragraph*{Proof of Theorem \ref{thm: consistency} (\romannumeral 1) } ~\\
Let $\mathcal{T}_2 = \cup_{g = G_1 + 1}^GA_g$ and define $\tilde{\boldsymbol{\theta}}_n = (\tilde{\theta}_{n1}, \dots, \tilde{\theta}_{nd})'$ with $\tilde{\theta}_{nk} = \hat{\theta}_{nk}$ if $k\notin \mathcal{T}_2$ and 0 otherwise. Then, the KKT condition for the Equation \eqref{appendix eq: objective function with neg log} implies that
\begin{equation}
    \frac{\partial L^c(\hat{{\theta}})}{\partial\theta} - \gamma\lambda_n\sum_{A_g \ni k}c_g\|\hat{{\theta}}_{nA_g}\|_1^{\gamma - 1}\text{sgn}(\hat{\theta}_{nk}) = 0,\quad \forall \hat{\theta}_{nk} \ne 0.
\end{equation}
Since $(\hat{\theta}_{nk} - \tilde{\theta}_{nk})\text{sgn}(\hat{\theta}_{nk}) = |\hat{\theta}_{nk}|I\{k \in \mathcal{T}_2\}$, we have

\begin{align}-\frac{\partial L^c(\hat{\theta}) }{\partial \theta} (\hat{{\theta}} - \tilde{{\theta}}) &= \sum_{k \in \mathcal{T}_2}|\hat{\theta}_{nk}|\gamma\lambda_n\sum_{A_g\ni k}c_g\|\hat{{\theta}}_{A_g}\|_1^{\gamma - 1} \nonumber\\
&= \gamma \lambda_n\sum_{g = 1}^Gc_g\|\hat{{\theta}}_{A_g}\|_1^{\gamma - 1} (\|\hat{{\theta}}_{nA_g}\|_1 - \|\tilde{{\theta}}_{nA_{g}}\|_1) \nonumber \\
&= \gamma \lambda_n\sum_{g = G_1 + 1}^Gc_g\|\hat{\theta}_{nA_g}\|_1^\gamma. \label{eq: appendix first deriv}
\end{align}
By the definition of $\hat{\theta}$, $Q(\hat{\theta}) \ge Q(\tilde{\theta})$. Since $\|\hat{\theta}_{nA_g}\|_1 = 0$ for $g > G_1$, and by the Equation \eqref{eq: appendix first deriv}, 

\begin{align*}
    &-\frac{1}{n}\frac{\partial L^c(\hat{\theta}) }{\partial \theta} (\hat{{\theta}} - \tilde{{\theta}}) + (1-\gamma)\frac{\lambda_n}{n}\sum_{g = G_1 + 1}^Gc_g\| \hat{{\theta}}_{nA_g}\|_1^\gamma\\
    &\le \frac{\lambda_n}{n}\sum_{g = 1}^G c_g\|\hat{{\theta}}_{nA_g}\|_1^\gamma- \frac{\lambda_n}{n}\sum_{g = 1}^Gc_g\|\tilde{{\theta}}_{nA_g}\|_1^\gamma\\
    &\le\frac{1}{n}[L^c(\tilde{{\theta}}) - L^c(\hat{{\theta}})]\\
    &=-\frac{1}{n}\frac{\partial \ell(\hat{\theta})}{\partial\theta}(\hat{{\theta}} - \tilde{{\theta}}) + \frac{1}{2}(\hat{{\theta}} - \tilde{{\theta}})^T[H({{\theta}}_0) + o_p(1)](\hat{{\theta}} - \tilde{{\theta}})  \\
    &\quad + o_p(\|\hat{{\theta}} - \tilde{{\theta}}\|_2^2)
\end{align*}
It follows that
\begin{align} 
&(1-\gamma)\frac{\lambda_n}{n}\sum_{g = G_1 + 1}^Gc_g\| \hat{{\theta}}_{nA_g}\|_1^\gamma \nonumber\\
&\le \frac{1}{2}(\hat{{\theta}} - \tilde{{\theta}})^T[H({{\theta}}_0) + o_p(1)](\hat{{\theta}} - \tilde{{\theta}}) \nonumber+ o_p(\|\hat{{\theta}} - \tilde{{\theta}}\|_2^2)\\
&= \frac{1}{2}(\hat{{\theta}} - \tilde{{\theta}})^T[H({{\theta}}_0) ](\hat{{\theta}} - \tilde{{\theta}}) + o_p(\|\hat{{\theta}} - \tilde{{\theta}}\|_2^2) \label{eq: appendix upper bound}
\end{align}
Since the first term of the Equation \eqref{eq: appendix upper bound} dominates the second term for large $n$, we have
\begin{align*}
    (1-\gamma)\frac{\lambda_n}{n}\sum_{g = G_1 + 1}^Jc_g\| \hat{{\theta}}_{nA_g}\|_1^\gamma
    &\le 2\rho_n^*\|\hat{{\theta}} - \tilde{{\theta}}\|_2^2 = 2\rho_n^*\|\hat{{\theta}}_{n\mathcal{T}_2}\|_2^2 \\
    &\le2\rho_n^*\|\hat{{\theta}} - {\theta}_0\|_2^2.
\end{align*}
It follows that
\begin{equation*}
    (1-\gamma)\lambda_n\sum_{g = G_1 + 1}^Gc_g\| \hat{{\theta}}_{nA_g}\|_1^\gamma\le 2n\rho_n^*{\|\hat{{\theta}}-{\theta}_0\|_2^2}= O_p(d_n\rho_n^*)
\end{equation*}
For the lower bound of $\sum_{g = G_1 + 1}^Gc_g\| \hat{{\theta}}_{nA_g}\|_1^\gamma$, since $c_j > 1$ by (C2), 
\begin{equation*}
    \sum_{g = G_1 + 1}^Gc_g\|\hat{{\theta}}_{nA_g}\|_1^\gamma \ge \Big(\sum_{g = G_1 + 1}^G\|\hat{{\theta}}_{nA_g}\|_1\Big)^\gamma \ge \|\hat{{\theta}}_{n\mathcal{T}_2}\|_1^\gamma \ge \|\hat{{\theta}}_{n\mathcal{T}_2}\|_2^\gamma.
\end{equation*}
Then, 
\begin{align*}
    (1-\gamma)\lambda_n \le2n\rho_n^*\|\hat{{\theta}}_{nB_2}\|_2^{2-\gamma} &= n\rho_n^*O_p(d_n/n)^{(2-\gamma)/2} \\
    &= \rho_n^*d_n^{1-\gamma/2}n^{\gamma/2}O_p(1)
\end{align*}
It follows that $\lambda_n/(n^{\gamma/2}\rho_n^*d_n^{1-\gamma/2})\le O_p(1)$. Since $\lambda_n/(n^{\gamma/2}\rho_n^*d_n^{1-\gamma/2})\rightarrow \infty$ by (C3), 
\begin{equation*}
    P\{ \|\hat{{\theta}}_{n\mathcal{T}_2}\|_2> 0\} \le P\Big\{\frac{\lambda_n}{n^{\gamma/2}\rho_n^*d_n^{1-\gamma/2}}\le O_p(1)\Big\}\rightarrow0.
\end{equation*}
\paragraph*{Proof of Theorem \ref{thm: consistency} (\romannumeral 2) } ~\\
Since $d_1, \theta_{0\mathcal{T}_1}$ are fixed, $\underset{g \le G_1}{\text{min}}\|{\theta}_{0A_g}\|_1^{1-\gamma} \equiv O(1)$ so that
\begin{equation*}
    \frac{\lambda_n^2}{n\rho_n}\sum_{g = 1}^{G_1}c_g^2\|{\theta}_{0A_g}\|_1^{2(\gamma - 1)}|A_g\cap \mathcal{T}_1| = O(1).
\end{equation*}
Thus, 
\begin{equation*}
    \|\hat{{\theta}}_{nB\mathcal{T}1} - {{\theta}}_{0\mathcal{T}_1}\|^2 = O_P(1/n), \quad \|\hat{{\theta}}_{n} - {{\theta}}_{0}\|^2 = O_P(1/n)
\end{equation*}
Let $h_n = n^{-1/2}$, and take $V_{1n}(u) = Q(\theta+0 + h_n(u', 0')' - Q(\theta_0)$, where 0 is a zero vector of dimension $|\mathcal{T}_2|$ and $u = (u_1, \dots, u_{d1})'$ is a $d_1$-dimensional constant vector. By part (\romannumeral 1) in Theorem \ref{thm: consistency}, with lage probability, $\hat{\theta} - \theta_0 = h_n(\hat{u}_n', 0')'$, and $\hat{u}_n = \text{argmin}\{V_{1n}(u): u \in \mathbb{R}^{d_1}\}$. Also, $V_{1n}$ can be rewritten as :
\begin{align*}
    V_{1n} &= \nabla L^c ({\theta}_0)^Th_n({u}',0') + \frac{1}{2}u^T\nabla^2 L^c({\theta}_0)u  + u'o_p(1)u\\
&\quad + \lambda_n\sum_{g = 1}^Gc_g\Big\{(\sum_{k\in A_g\cap \mathcal{T}_1}|\theta_{0k} + h_nu_k|)^\gamma - \|{\theta}_{0A_g}\|_1^\gamma\Big\}\\
&\equiv T_{1n}(u) + T_{2n}(u)
\end{align*}
By Central Limit Theorem, $n^{-1/2}\nabla L^c(\theta_0)\overset{d}{\rightarrow}W$, where $W = N(0, J(\theta_0))$ and $J(\theta) = \text{var}(\nabla L^c(\theta))$. In addition, $n^{-1}\nabla^2L^c(\theta_0) \overset{p}{\rightarrow}H(\theta_0)$. Thus, $T_{1n}(u) \overset{d}{\rightarrow}u'W + \frac{1}{2}u'H(\theta_0)u$. According to \cite{huang2009group}, 
\begin{align*}
    T_{2n}(u) \rightarrow\gamma\lambda_0\sum_{g = 1}^{G_1}c_g\|{\theta}_{0A_g}\|_1^{\gamma - 1}\sum_{k \in A_g\cap \mathcal{T}_1}\Big\{a_k\text{sgn}(\theta_{0k})I(\theta_{0k}\ne 0) \\+ |a_k|I(\theta_{0k} = 0)\Big\}.
\end{align*}
Therefore, $V_{1n}(u) \overset{d}{\rightarrow}V_1(u)$.

\section{Update Pseudo Response} \label{appendix: Update pseudo response}

This appendix provides the detailed proof of Proposition \ref{prop:mm} presented in Section \ref{subsec: algorithm} of the main text. We derive the update of the MM algorithm by maximizing the surrogate function. The original objective function is
\begin{equation*}
    f(Z_{ik,j}) = I(Y_{ik} = j)\bigg\{Z_{ik,j} - \log\bigg(\sum_{h=1}^{J_k} \exp(Z_{ik,h})\bigg)\bigg\}.
\end{equation*}
We can construct a quadratic minorizer using a second-order Taylor expansion around the current estimate $Z_{ik,j}^{(m)}$. The gradient vector $\nabla f(Z_{ik,j}^{(m)})$ has elements $\frac{\partial f}{\partial Z_{ik,h}} = y_{ik,h} - p_{ik,h}^{(m)}$, where $y_{ik,j} = I(Y_{ik} = j)$, $p_{ik,j}^{(m)}= P(Y_{ik} = j)$ calculated using $Z_{ik}^{(m)}$ (from now on abbreviated as $p_k$).  The Hessian matrix, $\nabla^2 f(Z_{ik,j}^{(m)})$, is bounded such that $\nabla^2 f(Z_{ik,j}^{(m)}) \succeq M_k$. The $(j,t)$ element of the hessian matrix is
\begin{align*}
    \frac{\partial^2f(Z_{ik,j})}{\partial Z_{ik,j}\partial Z_{ik,t}}
    &= \begin{cases}
        -\frac{\text{exp}(Z_{ik,j})\{1+\sum_{t= 2, t\ne j}\text{exp}(Z_{ik,t})\}}{\{1+\sum_{t = 2}^{J_k}\text{exp}(Z_{ik,t})\}^2}&\text{ if }j = t\\\frac{\text{exp}(Z_{ik,j} + Z_{ik,t})}{\{1+\sum_{t = 2}^{J_k}\text{exp}(Z_{ik,t})\}^2}&\text{ if }j \ne t
    \end{cases}\\
    &= \begin{cases}-p_{j}(1-p_{j})&\text{ if }j = t\\ p_{j}p_{t}&\text{ if } j \ne  t\end{cases}.
\end{align*} Thus,
\begin{equation*}
    \nabla^2 f(Z_{ik,j}) = \begin{pmatrix}-p_{2}(1-p_{2})&p_2p_3&\dots&p_2p_{J_k}\\\vdots &\vdots & &\vdots\\p_{J_k}p_2&p_{J_k}p_3&&-p_{J_k}(1-p_{J_k})\end{pmatrix}.
\end{equation*}
\paragraph*{When $J_k = 2$} ~\\
$\nabla^2 f(Z_{ik,j}) = -p_2(1-p_2) \ge -\frac{1}{4}$. Thus, $M_k = -\frac{1}{4}$ when $J_k = 2$.
\paragraph*{When $J_k = 3$}~\\ 
\begin{equation*}
    \nabla^2 f(Z_{ik,j}) = \begin{pmatrix}
        -p_2(1-p_2) & p_2p_3\\
        p_3p_2 & -p_3(1 - p_3)
    \end{pmatrix}.
\end{equation*}
We will use the eigenvalue of the Hessian matrix to obtain a lower bound. 
\begin{equation} \label{eq: det hessian}
    \text{det}\{\nabla^2f(Z_{ik,j}) - \lambda I\} = \{\lambda + p_2(1-p_2)\}\{\lambda + p_3(1-p_3)\} - p_2^2p_3^2.
\end{equation}
By setting Equation \eqref{eq: det hessian} as 0, 
\begin{align*}
    \lambda &= -\frac{p_2(1-p_2)  + p_3(1-p_3)}{2}\\
    &\;\pm \sqrt{\Big\{\frac{p_2(1-p_2)  + p_3(1-p_3)}{2}\Big\}^2 - p_2^2(1-p_2)^2p_3(1-p_3)^2 + p_2^2p_3^2}\\
    &\ge -\frac{p_2(1-p_2)  + p_3(1-p_3)}{2} \\
    &\quad- \sqrt{\Big\{\frac{p_2(1-p_2)  -p_3(1-p_3)}{2}\Big\}^2 + p_2^2p_3^2}\\
    &\ge -\frac{p_2(1-p_2)  + p_3(1-p_3)}{2} - \Big\{\frac{p_2(1-p_2) - p_3(1-p_3)}{2} + p_2p_3\Big\}\\
    &= -p_2(1-p_2) - p_2p_3\\
    &= -p_2(1-p_2+p_3)\\
    &\ge -p_2(2-2p_2)\\
    &\ge -\frac{1}{2}.
\end{align*}
Since the property of the characteristic equation of the matrix, the solution $\lambda$ is an eigenvalue, and the lower bound of the eigenvalue is $-\frac{1}{2}$. Thus, $M_k = -\frac{1}{2}$ when $J_k = 3$.
\paragraph*{When $J_k > 3$} ~\\
\begin{equation*}
    \nabla^2 f(Z_{ik,j}) = -\text{diag}(p) + pp^T,
\end{equation*}
where $p = (p_2, \dots, p_{J_k})$. Then, $\nabla^2 f(Z_{ik,j}) \ge -I$, so $M_k = -1$ for $J_k > 3$. 

Using this lower bound in the Taylor expansion provides the following minorizing surrogate function:
\begin{align*}
    f(Z_{ik,j}) &\ge I(Y_{ik} = j) \Bigg[Z_{ik,j}^{(m)} - \text{log }\Big\{\sum_{j = 1}^{J_k}\text{exp}(Z_{ik,j}^{(m)})\Big\}\Bigg] \\
    &\quad+ (Z_{ik,j} - Z_{ik,j}^{(m)})(y_{ik,j} - p_{ik,j}^{(m)}) + \frac{1}{2}M_k(Z_{ik,j} - Z_{ik,j}^{(m)})^2.
\end{align*}
The minimizer of the surrogate function is located at the critical point where its first derivative equals zero. The derivative of the surrogate function is
\begin{equation*}
    \frac{\partial g(Z_{ik,j}|Z_{ik,j}^{(m)})}{\partial Z_{ik,j}} = \frac{1}{M_k}(Z_{ik,j} - Z_{ik,j}^{(m)}) + y_{ik,j} - p_{ik,j}^{(m)}.
\end{equation*} 
Setting the derivative to zero and solving for $Z_{ik,j}$, we can get the analytical solution for the minimizer. The solution defines the update rule for the next iteration $(m+1)$:
\begin{equation*}
     \tilde{Z}_{ik,j}^{(m+1)} = Z_{ik,j}^{(m)} - \frac{1}{M_k} \left( y_{ik,j} - p_{ik,j}^{(m)} \right).
\end{equation*}

\end{document}